\documentclass[aps,nofootinbib,superscriptaddress,twocolumn,prd,10pt,floatfix]{revtex4-1}

\usepackage{amsmath}
\usepackage{mathrsfs}
\usepackage{graphicx}
\usepackage{amssymb}
\usepackage{comment}

\usepackage{hyperref}
\usepackage[dvipsnames]{xcolor}

\newcommand{\figref}[1]{Fig.~\ref{#1}}
\newcommand{\coH}{\mathcal{H}}

\begin{document}

\title{Phenomenology of Modified Gravity at Recombination}

\author{Meng-Xiang Lin}
\affiliation{Kavli Institute for Cosmological Physics, Department of Astronomy \& Astrophysics, Enrico Fermi Institute, The University of Chicago, Chicago, IL 60637, USA}
\author{Marco Raveri}
\affiliation{Kavli Institute for Cosmological Physics, Department of Astronomy \& Astrophysics, Enrico Fermi Institute, The University of Chicago, Chicago, IL 60637, USA}
\author{Wayne Hu}
\affiliation{Kavli Institute for Cosmological Physics, Department of Astronomy \& Astrophysics, Enrico Fermi Institute, The University of Chicago, Chicago, IL 60637, USA}

\begin{abstract}
We discuss the phenomenological imprints of modifications to gravity in the early universe with a specific focus on the time of recombination.
We derive several interesting results regarding the effect that such modifications have on cosmological observables, especially on the driving and phasing of acoustic oscillations, observed in the CMB   and BAO, as well as the weak gravitational lensing of the CMB and of galaxy shapes.
This widens the pool of measurements that can be used to test gravity with present and future surveys, in particular realizing the full constraining power of the structure of the primary peaks of the CMB spectrum.
We investigate whether such a phenomenology can relax tensions between cosmological measurements and find that a modification of the gravitational constant at recombination would help in reconciling measurements of the CMB with local measurements of the Hubble constant.
\end{abstract}
\maketitle
\section{Introduction} \label{sec:Intro}
Since the discovery of cosmic acceleration \cite{Riess:1998cb,Perlmutter:1998np}, understanding its physical origin has become one of the primary goals of experimental efforts to measure the cosmic microwave background (CMB) and the large scale structure (LSS) of the Universe.
In parallel it has been realized that the same measurements can be used to study gravity on cosmological scales including the possibility that modified gravity (MG) could explain cosmic acceleration
  (for reviews, see Refs.~\cite{Silvestri:2009hh,Clifton:2011jh,Joyce:2014kja}).
  
Both CMB measurements and LSS data, probing  the universe mainly at early and late times respectively, have proven to be extremely powerful in pursuing this program.   Current surveys already provide precision constraints \cite{Ade:2015rim,Joudaki:2016kym,Aghanim:2018eyx,Alam:2016hwk}
and future surveys, such as CMB-S4 \cite{Abazajian:2016yjj}, Euclid \cite{Laureijs:2011gra} and LSST \cite{Abell:2009aa}, are expected to greatly exceed their performance \cite{Zhao:2008bn,Zhao:2009fn,Hojjati:2011xd,Asaba:2013mxj} and test  General Relativity (GR) to unprecedented precision.

Most of the phenomenological effort in testing gravity on cosmological scales has been focused on constraining parameterized modifications to the Einstein equations relating the matter density contrast to the lensing and the Newtonian potentials \cite{Caldwell:2007cw,Amendola:2007rr,Hu:2007pj,Bertschinger:2008zb,Pogosian:2010tj,Pogosian:2016pwr} and has targeted the late times, during the epoch of cosmic acceleration \cite{Hojjati:2015ojt,Zhao:2017cud,Casas:2017eob,Peirone:2017ywi,Espejo:2018hxa}.

In this paper we discuss thoroughly  for the first time the phenomenology and observational imprints of such modifications to gravity at early times and especially at the time of recombination.
By enforcing the conservation of comoving curvature  on superhorizon scales, we determine the  initial conditions and evolution of perturbations in the radiation dominated epoch, relating their amplitude to inflationary perturbations.
We implement this approach in the Einstein-Boltzmann solver CAMB \cite{Lewis2000} extending the range of applicability of the MGCAMB code \cite{Zhao:2008bn,Hojjati:2011ix} to early times.

Considering a parametrization for deviations from GR that decouples early and late times, we discuss both analytically and numerically the behavior of linear perturbations at all epochs of the universe and their impact on cosmological observables.
In particular we focus on the MG imprint left on the acoustic peaks of the CMB power spectrum and 
baryon acoustic oscillations (BAO).  
We complete this analysis with a discussion of the MG effects on the clustering of LSS and on the lensing of the CMB and galaxies.
As a result we significantly enlarge the pool of measurements that can be used to test gravity, notably, with the inclusion of the full constraining power of CMB observations.

These extensions allow us to investigate whether MG effects at early times could explain existing discrepancies between cosmological datasets \cite{Raveri:2018wln} and in particular tensions between CMB measurements and low redshift probes (see also \cite{Lorenz:2017fgo} and references therein for related work on early dark energy).
We show that a different value for the effective gravitational constant at early times can partially relax tensions internal to the CMB dataset and between the CMB and local measurements of the Hubble constant and weak lensing of galaxies.
This is achieved by changing the CMB temperature, polarization and lensing predictions in a compatible manner due to the combined effect that MG has on the acoustic oscillations and on lensing and results in a preference for a larger gravitational constant at early times at greater than $98\%$ C.L.
We comment on why such a resolution is not possible with only late time modifications to gravity.

This paper is organized as follows. 
In Sec.~\ref{sec:Parametrization} we review the parametrized framework to modified gravity that we use and present initial conditions for superhorizon perturbations.
In Sec.~\ref{sec:PertEvo} we discuss analytically and numerically the behavior of cosmological perturbations at different epochs.
In Sec.~\ref{sec:Observables} we present the effect of such modifications on the acoustic peaks of the CMB and other cosmological observables.
In Sec.~\ref{sec:ParamEstimation} we review the tools that we use in practical tests of MG at early and late times.
In Sec.~\ref{sec:Results} we present the data constraints and discuss parameter degeneracies and the coordination of different physical effects to alleviate tensions.
We conclude in Sec.~\ref{sec:Conclusions}.
\vfill
\section{Parametrized deviations from $\Lambda$CDM} \label{sec:Parametrization}
In Sec.~\ref{sec:MGeqs} we review the general parameterized approach to modifications of gravity for linear perturbations  employed in Ref.~\cite{Hojjati:2011ix}, which assumes a metric theory of gravity with minimally coupled ordinary matter.
To extend this parametrization to early times, we derive the superhorizon solutions to the perturbation equations 
in Sec.~\ref{sec:INI} and relate the amplitude of perturbations above the horizon to the amplitude of curvature perturbations set by inflation.
We assume a given, but possibly modified, background Hubble expansion rate with a constant effective equation of state parameter.
These general relations are applied to specific cases that isolate the various aspects of the modifications in the following sections.

\subsection{Modified Gravity Equations}\label{sec:MGeqs}
In conformal Newtonian gauge, metric perturbations are specified by two gravitational potentials, the Newtonian potential $\Psi$ and the intrinsic spatial curvature potential $\Phi$, giving the line element in a spatially flat background:
\begin{equation}
ds^2 = a^2(\tau)[-(1+2\Psi)d\tau^2+(1-2\Phi)\delta_{ij}dx^idx^j] \,,
\end{equation}
where $a(\tau)$ is the scale factor as a function of conformal time $\tau$.
In addition the stress-energy tensor for the matter species that we consider is given at first order in perturbations by:
\begin{align}
T^0_{\,\,0} =\,& -\rho-\delta\rho \,, \nonumber\\
T^0_{\,\,j}  =\,& (\rho+P)v_j \,, \nonumber\\
T^i_{\,\,j}   =\,& (P+\delta P)\delta^i_{\;j} + \pi^i_{\;j} \,,
\end{align}
where $\rho$ and $\delta\rho$ are the average energy density and its perturbation, $P$ and $\delta P$ are the average pressure and its perturbation, $v_j$ is the fluid velocity and $\pi^i_{\;j}$ denotes the traceless ($ \pi^i_{\;i}=0$)
component of the stress-energy tensor perturbations. 
In Fourier space, the scalar component of the velocity can be expressed as its divergence $\theta \equiv ik^jv_j$ and the anisotropic stress by $(\rho+P)\sigma \equiv -(\hat{k}_i\hat{k}^j-\,\delta_{i}^{\;j}/3)\pi^i_{\;j}$.  
We assume that matter is still covariantly conserved in the metric so that its equations of motion follow 
\begin{align}
\delta \rho' + 3(\delta\rho+\delta P) &= -(\rho+P) \left( \frac{\theta}{\coH}  -3\Phi'\right), \\
[(\rho+P)\theta]' + 4(\rho+P)\theta &= \frac{k^2}{\coH}\Big[ \delta P - (\rho+P)\sigma \nonumber\\
&\quad + (\rho+P)\Psi \Big]  \,,
\label{eq:matterconservation}
\end{align}
where $'=d/d\ln a$ here and throughout.
The evolution of anisotropic stress $\sigma$ is given by the radiation Boltzmann equations and is also unmodified in form.  
Here  $\coH\equiv\dot{a}/a=aH$ is the conformal Hubble rate where dot denotes $d/d\tau$ here and throughout.

In GR, the Einstein equations determine the metric given the matter fields as:
\begin{align}
k^2\Phi =\,& -4\pi G a^2\Delta \rho \,, \\
k^2[\Phi-\Psi] =\,& 12\pi G a^2 (\rho+P)\sigma \,,
\end{align}
where $\Delta\rho \equiv \delta\rho+3\frac{\coH}{k^2}(\rho+P)\theta$ is the comoving-gauge density perturbation.  
Following Ref.~\cite{Hojjati:2011ix}, we modify these two Einstein equations with the addition of two free functions of time and scale, $\mu(a,k)$ and $\gamma(a,k)$, to phenomenologically parametrize deviations from GR in the two metric variables:
\begin{align}
k^2\Psi &= -4\pi \mu G a^2   [\Delta\rho+3(\rho+P)\sigma] \,, \label{eq:E1} \\
k^2[\Phi-\gamma \Psi]&  = 12\pi \mu G a^2 (\rho+P)\sigma \,, \label{eq:E2}
\end{align}
where $\mu$ parameterizes the effective gravitational constant as $\mu G$ while $\gamma$ encodes the ratio of the two potentials.
When both $\mu$ and $\gamma$ are equal to unity, the model reduces to GR and more generally we will refer to this parameterization as MG .

Notice that we parameterize the Poisson equation \eqref{eq:E1} with the lapse $\Psi$ rather than the curvature potential $\Phi$.
This convention highlights the fact that $\Psi$ enters directly into the dynamics of non-relativistic matter and makes it easier to implement the condition that its evolution given the background expansion depends only on $\Phi/\Psi$ above the horizon if the comoving curvature is conserved \cite{Hu:2007pj}.
In this sense, $\mu$ scales the overall gravitational effect of all matter while $\gamma$ encodes the difference between the gravitational effects on  non-relativistic and relativistic matter.

Also note that $\mu$ and $\gamma$ can in principle be arbitrary functions of time and scale allowing us to encompass the scalar sector of any metric theory of gravity where all matter species are minimally coupled to the metric (see, for example, Ref.~\cite{Ishak:2018his} for a recent review).

\subsection{Superhorizon Solutions and Initial Conditions}\label{sec:INI}
To utilize the MG parametrization described by Eqs.~(\ref{eq:E1}, \ref{eq:E2}) at early times, we have to derive initial conditions for perturbations when they are above the horizon.  
We present the main results in this section and further details on the specific case we consider in the next section can be found in Appendix \ref{sec:code}.

In the standard cosmological scenario, initial conditions for the perturbations are set by inflation.
Using the fact that comoving gauge curvature is still conserved as $k/\coH \rightarrow 0$ (see Appendix \ref{app:CoR}) we can compute the initial conditions at the time when inflation ends and set them for perturbations later on, before the given mode re-enters the horizon.
 
We can thus derive the superhorizon solutions and initial conditions for perturbations in any gauge assuming some background Friedmann-Robertson-Walker expansion.  
If ${\cal R}$ is constant, Ref.~\cite{Hu:2016wfa} shows, using a separate universe argument,  that in any metric theory of gravity
\footnote{In comoving gauge of the metric as defined in Ref.~\cite{Hu:2016wfa}, Eq.~(\ref{eq:Rgaugetrans}) is exact and coincides with comoving gauge as defined by the matter velocity as $k/\coH\rightarrow 0$ (see Eq.~\ref{eq:Reqtheta}).}
with a spatially flat background
\begin{equation}
{\cal R} \approx -\Phi +\frac{H}{H'} (\Psi +\Phi').
\label{eq:Rgaugetrans}
\end{equation}
  This equation has the formal solution
\begin{equation}
\Phi = - \left( 1 - \frac{H}{g} \int d\ln a \frac{g'}{H} \right){\cal R} + C\frac{H}{g},
\label{eq:formal}
\end{equation}
where the integrating factor $g=e^{\int (\Psi/\Phi) d\ln a} $ and $C$ is an integration constant for what is
generally a decaying mode.

Assuming that the background expansion has a  Hubble rate $H \propto a^{-3(1+w)/2}$, we can solve this equation for $\Psi$ and $\Phi$.   
If the Friedmann equation itself is modified, $w$ simply parameterizes the expansion history and is not necessarily the equation of state parameter of the matter.
For the case where $\Phi/\Psi$ does not vary on the Hubble time scale or faster, which is true for most models during an epoch when $w=\,$const.,
the growing mode of Eq.~(\ref{eq:formal}) is solved by constant $\Phi$ and $\Psi$.    
Assuming that the anisotropic stress of the matter is dominated by neutrinos, we can integrate their equation of motion to find these constants.   
Since this equation is not modified in form, the result is the same as in GR
\begin{equation}
\sigma_\nu = \frac{8}{45}\frac{1}{1+ 4 w+ 3 w^2} \left( \frac{k}{\coH} \right)^2 \Psi,
\end{equation}
so that
\begin{equation}
\Phi -\gamma\Psi = \frac{16}{15} \frac{\mu R_\nu}{1+ 4 w+ 3 w^2} \Psi,
\end{equation}
where  $R_\nu = 8\pi G \rho_\nu/ 3 H^2$.  
Combining this with Eq.~(\ref{eq:Rgaugetrans}), we obtain
\begin{align}
\Psi &= -\frac{ 15 (1+ 4 w + 3 w^2)}{10+16\mu R_\nu + 30 w + 15 \gamma(1+ 4 w+ 3 w^2)}{\cal R}, \nonumber\\
\Phi &= -\frac{16\mu R_\nu + 15 \gamma (1+4w+3w^2)}{10+16\mu R_\nu + 30 w + 15 \gamma(1+ 4 w+ 3 w^2)}{\cal R}.
\end{align}
Our starting assumption that $\Phi/\Psi \approx\,$const.\ implicitly requires constant $\gamma$ and $\mu R_\nu$, but does not place other restrictions
on whether the Friedmann equation is itself modified. 
This suffices for our purposes since $R_\nu$ is constant for a radiation dominated expansion with $w=1/3$ and the neutrino anisotropic stress becomes negligible in other limits.
For the more general case, one can solve for the evolution of $\Phi$ and $\Psi$ given the evolution of $\gamma$ and $\mu R_\nu$ by supplementing Eqs.~(\ref{eq:E2}) and (\ref{eq:Rgaugetrans}) with their time derivatives.

\section{Perturbation Evolution}\label{sec:PertEvo}

From this point forward, we focus on an illustrative, but scale independent, parameterization of the parameters $\mu(a)$ and $\gamma(a)$, which isolates either early time or late time modifications to gravity, with the former being new to this work.  
This is achieved by using a smoothed step functional form for these functions, as discussed in Sec.~\ref{sec:step}.
Furthermore, we assume an unmodified $\Lambda$CDM background to isolate the effect of the modifications to linear perturbation theory.    
We present  analytic results for perturbation evolution in Sec.~\ref{sec:analytic} and  numerical results in Sec.~\ref{sec:FullNumSolution}. 
Most of the analytic results can be easily extended to the scale dependent case since each $k$-mode evolves independently in linear theory.

\subsection{Step Parameterization}\label{sec:step}

In order to separate early time and late time effects on perturbations, we shall consider a phenomenological parametrization that models a transition between these two regimes.
Furthermore, to isolate the effects of the modified perturbation equations from the influence of the background expansion, we assume an unmodified $\Lambda$CDM expansion history from this point forward.

In particular we parameterize $\mu$ and $\gamma$ as step-functions in e-folds $N\equiv\ln{a}$ with the following smooth step-like form:
\begin{equation}
f(x) = \frac{ f_0+f_\infty}{2} - \frac{ f_0-f_\infty}{2} \frac{x}{\sqrt{1+x^2} } \,,
\end{equation}
where $x = (N-N_T)/\Delta_T$ and $f \in \{\mu,\gamma\}$. Here we have four parameters: $f_0$ and $f_\infty$ are the values of the quantity today and at early times respectively, $N_T\equiv\ln{a_T}$ denotes the e-folds of transition between the two regimes, and $\Delta_T$ is the e-fold width of the transition. 
In this paper, we set $N_T=-3.4$, corresponding to $z \sim 30$ as it is  approximately the median of the $\sim 6$ e-folds between recombination and today. 
As such, the transition happens at a time that is well after recombination, before the late time accelerated expansion and beyond the reach of the next generation of large scale structure surveys.
We further choose $\Delta_T=1$ to avoid a sharp transition which would introduce spurious effects on the CMB power spectrum.  

We  test the stability of our results to the choice of  transition width in Appendix \ref{sec:TransitionWidth}. 
The results depend only weakly on $\Delta_T$ if it is around $1$, and become sensitive to  $\Delta_T$ when it is much larger since the transition would affect physical processes around recombination.
When $\Delta_T$ is very small the results converge to a unique answer, but produce spurious effects on the CMB power spectrum which would not be present if the transition occurred on the Hubble time scale or greater.

Thus, in addition to $\Lambda$CDM parameters, we have four additional free parameters: $\mu_0$, $\mu_\infty$, $\gamma_0$, $\gamma_\infty$. 
Note that for more realistic gravity models we usually expect the two functions $\mu$ and $\gamma$ to be scale dependent.

\subsection{Analytic Results}\label{sec:analytic}

In this section we show some analytic results that help interpret the novel features of perturbation evolution in MG at early times.  
We discuss the limiting cases of superhorizon and subhorizon evolution in the various epochs of a $\Lambda$CDM background expansion.   
Using the step parameterization, we isolate the impact of a nearly constant $\mu$ and $\gamma$ at early and late times.

\subsubsection{Superhorizon Solutions}\label{sec:superhorizon}

On superhorizon scales, our general derivation in the previous section applies before and after the step.
In particular, given our assumption that the background expansion history is unmodified, we have in the 
radiation dominated epoch
	\begin{align}\label{eq:iniNewtonian}
	\Psi =\,& -\frac{10}{10\gamma+5+4\mu R_\nu} \mathcal{R} \,, \nonumber\\
	\Psi+\Phi=\,&  -\left(1+\frac{5}{10\gamma+5+4\mu R_\nu}\right)\mathcal{R} \,.
	\end{align}
where $(\Psi+\Phi)/2$ is the Weyl potential that enters into gravitational lensing and the integrated Sachs-Wolfe
effect in the CMB.
While both $\gamma$ and $\mu$ larger than one would result in a smaller amplitude of the initial gravitational and Weyl potentials, their quantitative effect is different.
In particular, changing $\mu$ results in a smaller change in the gravitational potentials, with respect to $\gamma$, since it is multiplied by the neutrino fractional energy density $R_\nu$. 
This is a consequence of the fact that, given the background expansion, superhorizon potentials depend only on $\Phi/\Psi$  which itself is determined by $\gamma$ alone in the
absence of anisotropic stress.
This behavior of the Weyl potential plays an important role in understanding the modified CMB power spectrum that we shall discuss in Sec. \ref{sec:CMB}. 

For modes that remain outside the horizon after radiation domination, when $R_\nu \ll 1$, the solutions well before or after the step reduce to 
\begin{eqnarray}
\Psi &=& -\frac{ 3(1+w)}{2 + 3\gamma(1+w)}\mathcal{R}
\,, \label{eq:super-MD_Psi} \\
\Psi+\Phi &=& -\frac{3 (1+w) }{2+ 3\gamma(1+w) } (1+\gamma)\mathcal{R}
\,, \label{eq:super-MD_Weyl}
\end{eqnarray}
and in particular in the matter dominated limit $w=0$ and these relations further simplify.
Notice that after radiation domination the superhorizon solution does not depend on $\mu$ but only $\gamma$.
During the recent acceleration epoch, well after the step, $w$ is not constant and so these solutions do not strictly apply 
but since $\Phi/\Psi = \gamma$, Eq.~(\ref{eq:formal}) implies
\begin{align}
\Psi & = - \left( 1 - \frac{H}{\gamma a^{1/\gamma}} \int d\ln a \frac{a^{1/\gamma}}{H} \right) \frac{\cal R}{\gamma}, \nonumber\\
\Psi+\Phi &=  \left( 1+\gamma \right)\Psi.
\label{eq:formalacc}
\end{align}
This integral relation can be expressed in terms of the hypergeometric function for the $\Lambda$CDM expansion history
which shows that  Eq.~(\ref{eq:super-MD_Psi},\ref{eq:super-MD_Weyl}) qualitatively describe the transition to $\Lambda$ domination but predicts it to be more rapid.
A larger $\gamma$ also causes a slower
decay of the potential during the acceleration epoch.

\begin{figure*}[!ht]
\centering
\includegraphics[width=\textwidth]{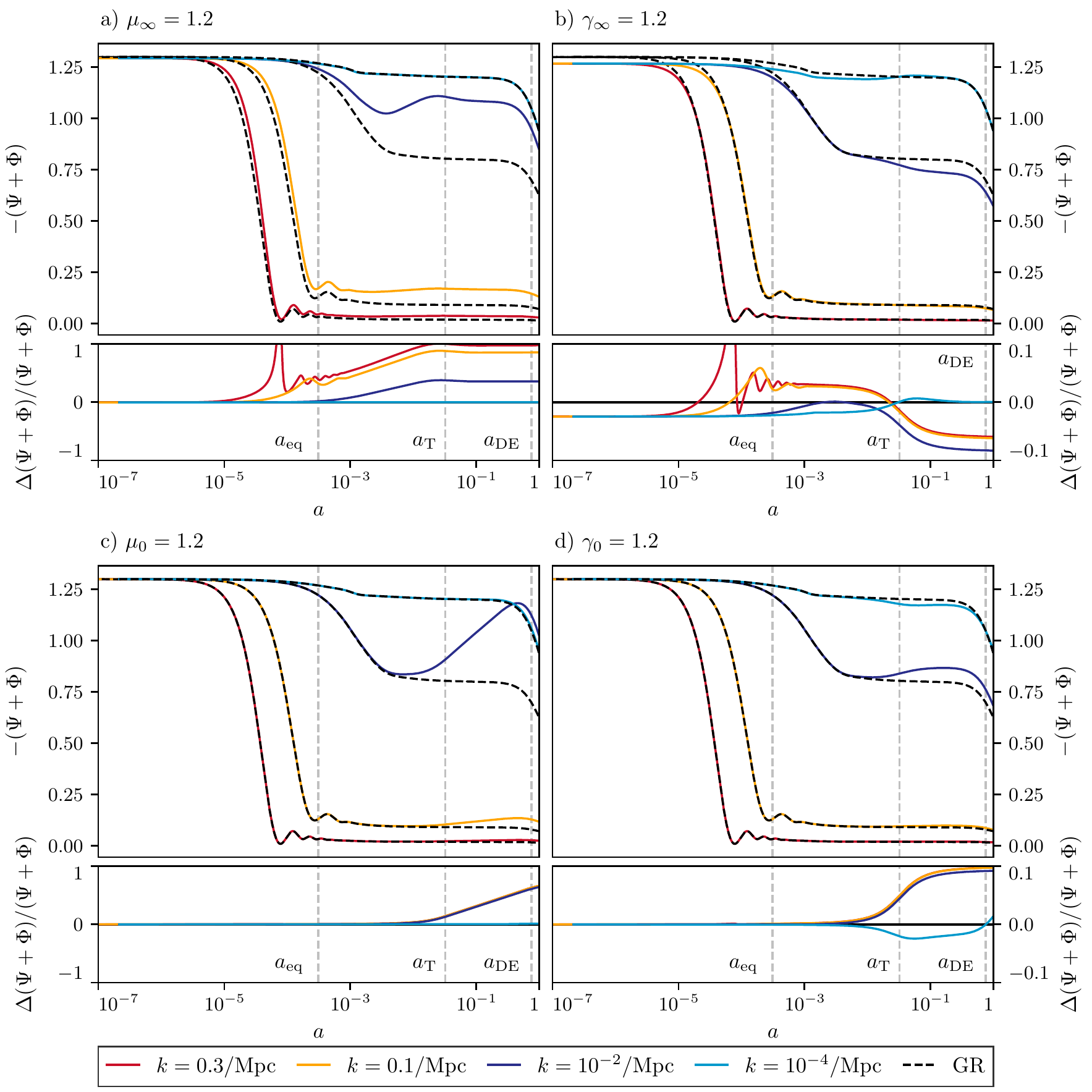}
\caption{\label{fig:Weyl_all}
The comparison of the Weyl potential evolution between our MG example models and GR. The four panels represent models with $\mu_\infty=1.2$, $\gamma_\infty=1.2$, $\mu_0=1.2$ and $\gamma_0=1.2$ respectively. The three vertical dashed lines indicate, from left to right respectively, matter-radiation equality, the transition of the MG parameters (here $z=30$, see definition in Sec.~\ref{sec:step}), and $\Lambda$-matter equality.
Before horizon crossing, $\mu$ has a limited effect on the evolution of Weyl potential due to the small fraction of neutrino energy density, while a larger $\gamma$ decreases its amplitude in both radiation and matter dominated epochs and slows the potential decay in the acceleration epoch. 
When crossing the horizon during radiation epoch, a larger $\mu$ delays the potential decay significantly while the same change in $\gamma$ leads to a small effect.
After horizon crossing, a larger $\mu$ increases the amplitude of the potential due to a larger effective gravitational constant $\mu G$, while $\gamma$ does not affect the subhorizon evolution. 
For details of the early and late time behaviors and the effect of the transition, see the discussion in Sec.~\ref{sec:FullNumSolution}.
}
\end{figure*}
\begin{figure*}[!ht]
\centering
\includegraphics[width=\textwidth]{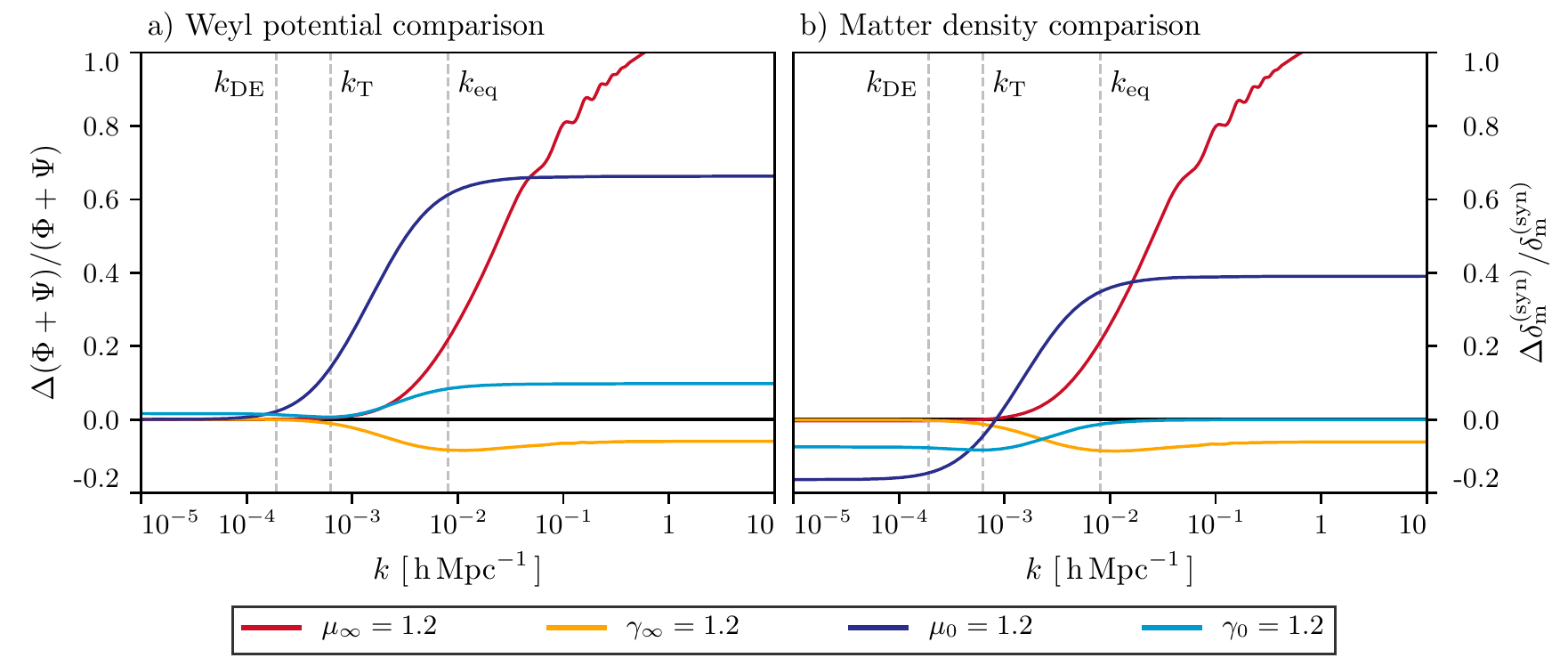}
\caption{
\label{fig:transfer_all}
The fractional change in transfer functions relative to their GR values at redshift $z=0$ due to MG of (a)  the Weyl potential and    (b) the synchronous gauge matter density perturbations. 
Different colors represent different example models as in \figref{fig:Weyl_all}, as shown in legend.
The vertical lines show the scales, $(k_{\rm DE},k_{\rm T},k_{\rm eq})$, corresponding to the modes crossing the horizon at $\Lambda$-matter equality, transition in the MG functions, and matter-radiation equality respectively.
}
\end{figure*}
\subsubsection{Subhorizon Solutions}\label{sec:subhorizon}
In the subhorizon regime, the evolution of the matter perturbations determines the evolution of the potentials and vice versa.  
Since the equations of motion of matter given the metric are not affected by MG, we first examine its behavior given the metric.  
		
Deep in the radiation dominated epoch, the baryon density is negligible and the photon density, or monopole perturbation $\Theta_0=\frac{1}{4}\delta_\gamma$, can be formally expressed as (see Eq.~D-6 in Ref.~\cite{Hu1995}\footnote{Note that $\Phi$ in Ref.~\cite{Hu1995} is $-\Phi$ in this paper and $V=\theta/k$, $p\Pi = 3(\rho+P)\sigma/2$.}):
\begin{eqnarray}\label{eq:Theta0-Phi}
[\Theta_0-\Phi](\tau) & =&  [\Theta_0-\Phi](0)\cos (k c_s \tau)\\ 
&& - \frac{k}{\sqrt{3}}\int_0^\tau d\tau^\prime [\Phi+\Psi](\tau^\prime)\sin[ k c_s (\tau-\tau') ] \,, \nonumber
\end{eqnarray}
where $c_s=1/\sqrt{3}$.
Since the radiation density fluctuation dominates the source of the Poisson equation, once the photons enter into acoustic oscillations around sound horizon crossing, the Weyl potential decays to zero.   
This decay also provides a source to the acoustic oscillations through the integral term in Eq.~(\ref{eq:Theta0-Phi}) which we refer to as the radiation driving effect.   
This extra source can carry a phase shift if the timing of the decay is modified. 
This same effect causes the well-known phase shift due to freestreaming neutrinos (e.g.~\cite{Baumann:2015rya}).
		
As we shall see, the phenomenology of acoustic oscillations is then determined by two pieces: the initial superhorizon conditions for $\Theta_0-\Phi$, the Weyl potential and the modification of the timing of the decay of the latter. 
Since $\Theta_0(0)  = -\Psi(0)/2$, MG does not change the initial value of $[\Theta_0-\Phi](0)=\mathcal{R}$, so the change of this solution comes from the integral of the Weyl potential. 
Since the Weyl potential always decays to zero after horizon crossing in the radiation dominated epoch, the amplitude of the driving effect depends on its initial value.
We see from Eq.~\eqref{eq:iniNewtonian} that a larger $\gamma$ gives a lower value of $|\Phi+\Psi|(0)$, which decreases the driving effect. 
The impact of $\mu$ on the amplitude is smaller but it does influence the timing of the decay.   
A larger $\mu$ provides a larger source to the Weyl potential and delays the decay.  
This then produces a phase shift in the acoustic oscillations as we shall see in Sec.~\ref{sec:CMB}.  
		
Cold dark matter density perturbations $\delta_c$ evolve according to:
\begin{equation}\label{eq:cdmevoS}
\ddot{\delta}_c + \frac{\dot{a}}{a}\dot{\delta}_c = -k^2\Psi-3\ddot{\Phi} \,.
\end{equation}
In the radiation dominated epoch, we can treat the right hand side as an external driving force $S(k,\tau)=-k^2\Psi-3\ddot{\Phi}$. 
Given that the potentials decay at horizon crossing, as discussed above, well after horizon crossing $\delta_c$ settles into a logarithmic growing mode \citep{Hu1996}:
\begin{equation}\label{eq:delta-RD}
\delta_c(k,\tau) = -A\Psi(k,0)\ln(Bk\tau) \,,
\end{equation}						
where $A$ and $B$ are constants that can be determined from
\begin{eqnarray}
A&=& -\frac{1}{\Psi(k,0) }\int_0^\infty d\tau S(k,\tau)\tau \,, \\
A\ln B &=& \frac{3}{2} + \frac{1}{\Psi(k,0) }\int_0^\infty d\tau S(k,\tau)\tau\ln(k\tau) \,.\nonumber
\end{eqnarray}			
The MG effect therefore again comes from the initial conditions and the timing of the decay. 
Recall also that $\tau \propto a$ in the radiation dominated epoch. 
Even though $A$ and $B$ themselves have no $k$-dependence deep in the radiation dominated regime, a change
in $B$ alters  the  transition in $k$ between the constant and $\ln (k\tau)$ terms in $\delta_c$.   A change
in $B$ occurs when the epoch of potential decay is shifted.
This is especially pronounced for $\mu$ whereas $\gamma$ mainly changes the overall amplitude $A$.

After the Universe becomes matter dominated, the self gravity of matter causes its density fluctuation to grow due to $k^2\Psi$ whereas $\ddot \Phi$ remains negligible.  
After recombination, the baryon density fluctuation also obeys Eq.~(\ref{eq:cdmevoS}) and so the combined baryon and cold dark matter component is
\begin{equation}\label{eq:cdmevo}
\delta_m'' + \left( 2 + \frac{H'}{H}\right) \delta_m' + \frac{k^2}{\coH^2}\Psi =0,
\end{equation}
as usual.
The MG influence comes from the Poisson equation (\ref{eq:E1}) for $\Psi$ and involves $\mu$ whereas $\gamma$ drops out of the equations.
A larger $\mu$ increases the amplitudes of the potentials due to the larger effective gravitational constant and therefore enhances the growth.   
During the matter dominated epoch, neglecting the effect of massive neutrinos for simplicity, we have
\begin{equation}\label{eq:cdmevo-MD}
\delta_m'' +\frac{1}{2} \delta_m' - \frac{3}{2} \mu \delta_m = 0 
\end{equation}
and therefore for the growing mode
\begin{equation}\label{eq:deltac-MD}
\delta_m \propto a^{\frac{\sqrt{24\mu+1}-1}{4}} \,.
\end{equation}
When $\mu=1$, it reduces to the standard $\delta_m\propto a$ solution. 
When $\mu$ deviates from unity, the matter perturbation increases as $\mu$ increases.  
Massive neutrinos slow the growth rate in the same way below their freestreaming scale by acting in the opposite sense 
as a component that modifies the background expansion but does not contribute to the perturbations.

In the acceleration epoch, to good approximation $\delta_m \propto D^{\frac{\sqrt{24\mu+1}-1}{4}}$
where $D$ is the GR linear growth function of $\Lambda$CDM \cite{Hu:1997vi}.

\subsection{Numerical Results}\label{sec:FullNumSolution}

In this section we complement our analytical analysis with results from numerical integration using the modified Einstein-Boltzmann code described in Appendix \ref{sec:code}.

In \figref{fig:Weyl_all}, we show the numerical solution for the Weyl potential at different scales while in \figref{fig:transfer_all} we show the relative comparison of the transfer functions of the Weyl potential and synchronous gauge matter density perturbations at redshift zero.
We choose four example models defined by $\mu_\infty=1.2$, $\gamma_\infty=1.2$, $\mu_0=1.2$, $\gamma_0=1.2$ while keeping fixed all other cosmological parameters.
We refer to the scale factors at matter-radiation equality, MG parameter transition, and matter-dark energy equality as, respectively, $a_{\rm eq}$, $a_{\rm T}$, and $a_{\rm DE}$. The wavenumbers of the modes that enter the horizon at the corresponding times, which are determined by $k_i=\coH(a_i)$, are referred to as $k_{\rm eq}$, $k_{\rm T}$, and $k_{\rm DE}$. 

As we can see from \figref{fig:Weyl_all}, before horizon crossing, as expected from the analytic results, $\mu$ has a very limited impact on the evolution of the Weyl potential, while a larger $\gamma$ decreases its amplitude in both radiation dominated and matter dominated epochs and slows the potential decay in the acceleration epoch. 

At horizon crossing, for modes that cross during radiation domination, an increase in $\mu$ delays the decay of the Weyl potential while the same change in $\gamma$ leads to a much smaller effect. 
As we see in Eq.~(\ref{eq:Theta0-Phi}), this implies that $\mu$ being different from its GR value through recombination changes the phase of acoustic oscillations in the CMB. 
Modes that enter the horizon before matter-radiation equality ({\it i.e.}~$k>k_{\rm eq}$) still grow logarithmically but if $\mu_{\infty}\neq 1$, the change in the decay epoch also  changes  the constant vs.~logarithmic coefficients that results in an enhancement that grows  with $k$, as shown in   \figref{fig:Weyl_all}a. 
$\gamma$ mainly changes the overall amplitude of the decay and therefore leads to much less scale dependence.

After horizon crossing, during the matter and acceleration epochs, a larger $\mu$ increases the growth rate of perturbations due to a larger effective gravitational constant $\mu G$.
The relative change in growth is scale independent during the  matter and acceleration epochs, since we assumed a scale independent parametrization for $\mu$.

We now comment on the behavior of perturbations when crossing the transition of the MG functions at $z=30$. 
For subhorizon modes, the time derivative term $\Phi'$ in the continuity equation (\ref{eq:matterconservation}) becomes negligible, so that the density perturbations remain continuous even when $\gamma$ or $\mu$ change rapidly.  
From the Poisson equation (\ref{eq:E1}) we know that $\Psi\propto\mu$, and from Eq. \eqref{eq:E2} we have $\Phi=\gamma\Psi\propto\mu\gamma$ and hence $(\Psi+\Phi)\propto\mu(1+\gamma)$. 
Therefore, a transition in $\mu(1+\gamma)$ results in a transition in the Weyl potential that can be clearly seen for $\gamma$ on subhorizon scales in \figref{fig:Weyl_all}.
This feature is present but hidden in the $\mu$ case due to the change in the growth of density perturbations that overcomes this effect.

For superhorizon modes in the matter dominated limit, $\mu$ is irrelevant for the two potentials and remains so during the transition, we thus see zero impact in the $\mu_\infty$ and $\mu_0$ example models. 
For $\gamma$, on the other hand, the potentials settle on the predictions from Eq.~(\ref{eq:super-MD_Weyl}) before and after the transitions since the conservation of comoving curvature ensures that there is no memory of the transition once it is complete.

Combined these changes in growth lead to features in the transfer functions at $z=0$ displayed in  \figref{fig:transfer_all}.
At scales that are superhorizon at $k_{\rm DE}$, only $\gamma_0$ shows a relative deviation from GR, in the Weyl potential.
For a  transition from an unmodified  to a modified gravity model, $\mu=1 \rightarrow 1.2$, the scale independent change in growth appears as a 
scale independent increase in the transfer function at $k\gg k_{\rm T}$. 
On the other hand with a transition from a modified to an unmodified gravity model $\mu=1.2 \rightarrow 1$, this matter dominated growth only happens between horizon crossing
and $a_T$ so that its full effect occurs for $k\gg k_{\rm eq}$.  In addition there is also a scale-dependent component due to the 
modified logarithmic growth during radiation domination.

For both $\gamma_0$ and $\gamma_\infty$, the transition produces a step in the Weyl potential for modes $k>k_{\rm T}$ 
and hence a step in the Weyl transfer function.   For $\gamma_\infty$ there is an additional  change in the logarithmic growth during radiation domination that partially compensates for this step.

This discussion is tightly connected with the behavior of matter density perturbations whose scale dependence at $z=0$ can be seen in \figref{fig:transfer_all}.
There we show the relative comparison of the synchronous gauge matter density perturbations with respect to their GR behavior.

For both $\gamma_\infty$ and $\mu_\infty$ the behavior is very close to that of the Weyl potential (for more detail, see above discussion).
For the late time models, $\mu_0$ and $\gamma_0$, some differences appear. 
In the $\mu_0$ model, the amplitude on superhorizon scales decreases because $\mu$ does not affect the evolution of $\Psi$ at such scales, thus $\delta_c\propto 1/\mu$. 
In the $\gamma_0$ model, the amplitude on superhorizon scales decreases because its impact on $\Psi$ and hence $\delta_c$ differs from $\Phi+\Psi$, see Eq.~(\ref{eq:super-MD_Psi}).
For the modes that enter the horizon well before $z=30$, the amplitude remains unchanged because $\gamma$ does not affect the subhorizon matter perturbation evolution.

\begin{figure*}[!ht]
\centering
\includegraphics[width=\textwidth]{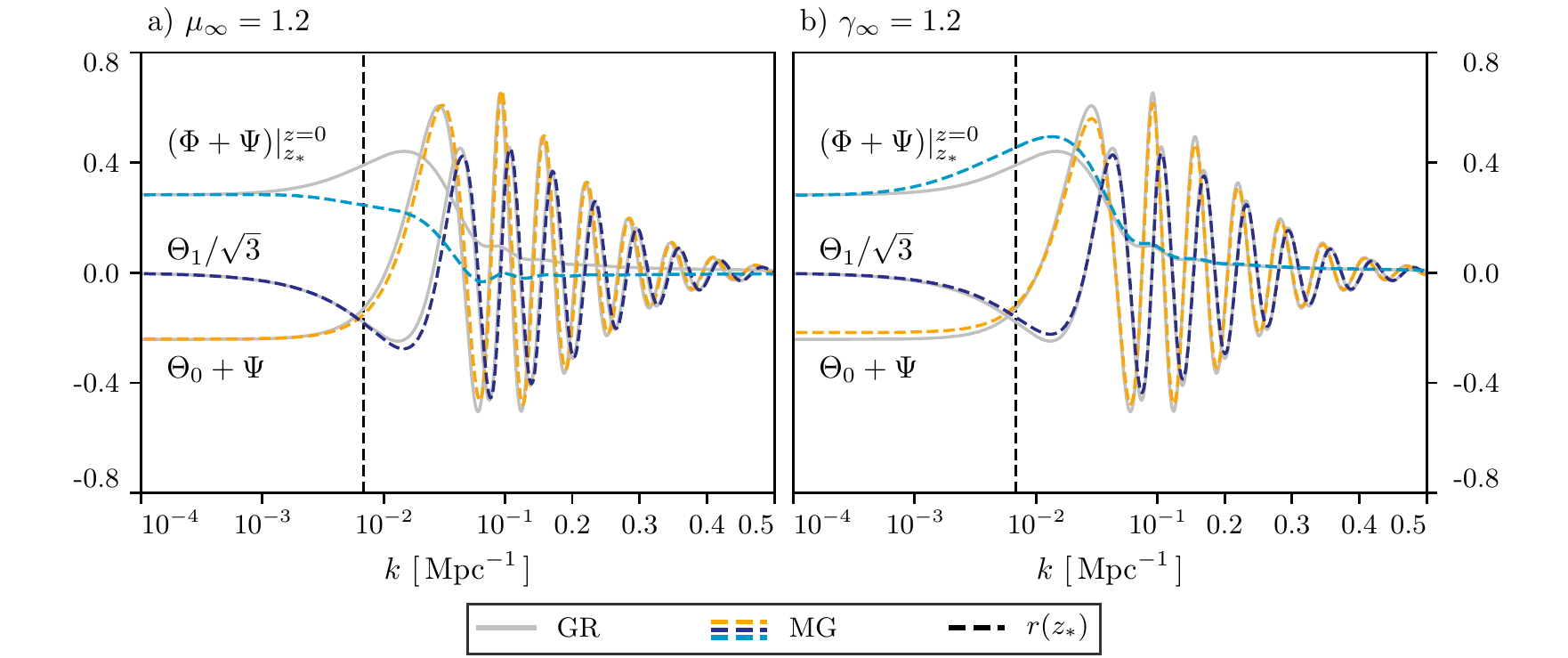}
\caption{\label{fig:CMBkspace}
The CMB anisotropy source functions in $k$-space in units of amplitude of primordial comoving curvature perturbation in two MG example models with $\mu_\infty=1.2$ and $\gamma_\infty=1.2$ and GR.
Different lines correspond to different physical effects and models, as shown in figure and legend. 
The vertical dashed line shows mode that crosses the horizon at recombination $(z_*)$.
}
\end{figure*}	
\begin{figure*}[!ht]
\centering
\includegraphics[width=\textwidth]{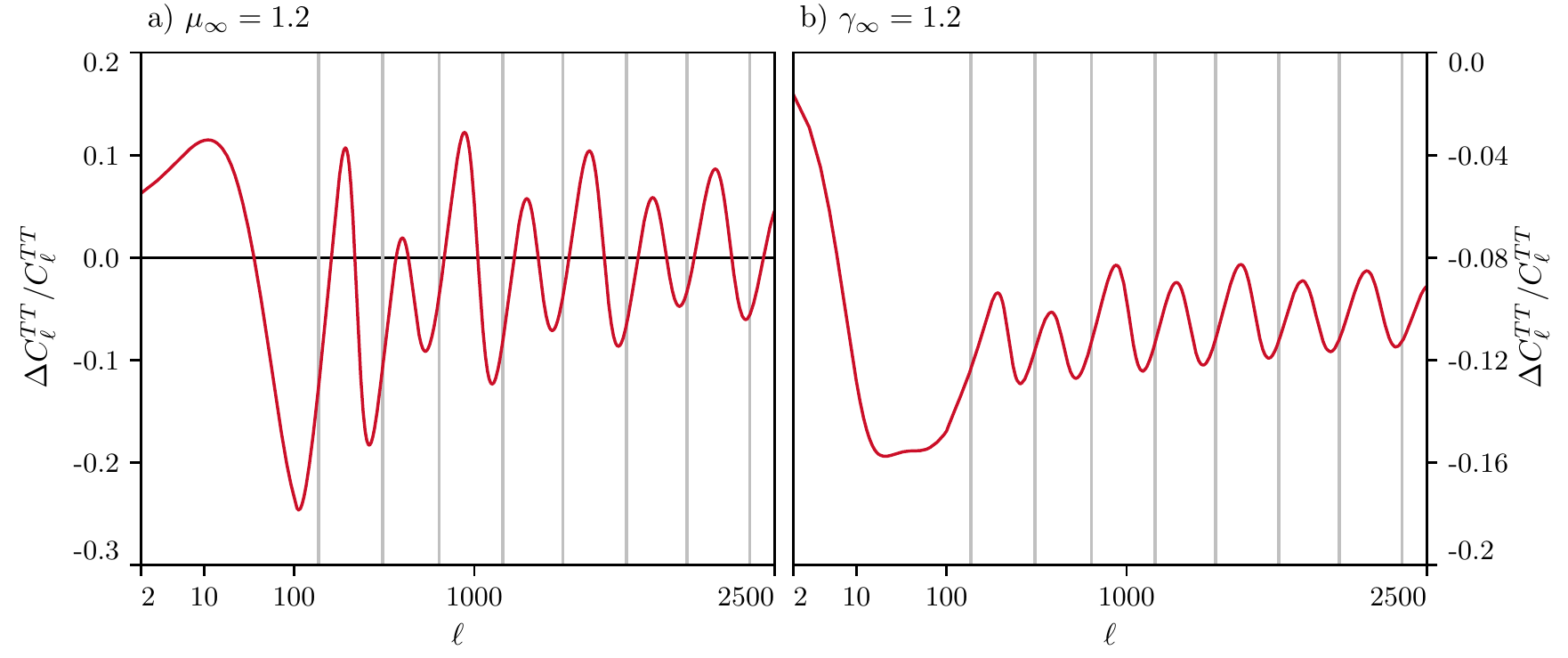}
\caption{\label{fig:CMBlspace}
The fractional change in the unlensed CMB temperature spectrum in two MG example models with $\mu_\infty=1.2$ and $\gamma_\infty=1.2$ relative to the GR spectrum. 
The vertical solid lines indicate the angular position of the GR peaks of the unlensed CMB spectrum.  Notice
that the variations are mainly out of phase with the peaks.  }
\end{figure*}	
\section{Impact on Observables}\label{sec:Observables}
In this section we study the impact of MG on cosmological observables, including the CMB power spectrum, the BAO scale, gravitational lensing and the matter power spectrum.

\subsection{CMB temperature power spectrum}\label{sec:CMB}

As we have seen in the previous sections, early time MG affects both initial conditions and the evolution of perturbations.
Therefore we expect a change in the physics of photon-baryon acoustic oscillations that will have a significant impact on the CMB power spectra. 
Since the Boltzmann equations remain unchanged in form, if we treat the gravitational potential as an external driving force for the photon-baryon fluid, the impact of MG can be understood from these changes in the potential evolution in the same way as  in GR \citep{Hu1995}.

To aid our interpretation of these effects, we start with the impact on the sources of CMB anisotropy in $k$-space in \figref{fig:CMBkspace}. There we show  $\Theta_0+\Psi$, the monopole corrected for the ordinary Sachs-Wolfe effect, and the dipole $\Theta_1$ at the redshift of recombination $z_*$. We also show the difference in Weyl potential between recombination and today $(\Phi+\Psi)|^{z=0}_{z_*}$ as a proxy for the integrated Sachs-Wolfe (ISW) effect, including its early time contribution.  
\figref{fig:CMBkspace} shows these three quantities in GR and in two example models with $\mu_\infty=1.2$ and $\gamma_\infty=1.2$ respectively. 
In addition we can see in \figref{fig:CMBlspace} the un-lensed scalar part of the CMB temperature spectrum for the two example models compared to the GR scalar spectrum in the harmonic domain.

As we can see $\mu_\infty$ induces a significant phase shift of both the temperature monopole and dipole. 
This phase shift comes from the shift in the epoch that the Weyl potential drives the oscillations as discussed in the previous section, see Eq.~(\ref{eq:Theta0-Phi}), and shown in the top left panel of \figref{fig:Weyl_all}.
This is the leading effect that we see, with respect to GR, in \figref{fig:CMBlspace}(a).

In the angular power spectra,  this phase shift corresponds to a shift of the scale of the acoustic peaks to higher multipole
with $\mu_\infty =1.2$ vs.~$0$ of $\Delta \ell \simeq 18$, where the exact number is calibrated on the third peak.  In \figref{fig:CMBlspace} this is visible as oscillatory fractional changes to the spectrum that are out of phase with the peaks themselves (vertical lines).   This should be contrasted
with a change in the fundamental angular scale of the acoustic peaks  $\theta_s$ which causes a shift $\Delta\ell \simeq -\ell \Delta \theta_s/\theta_s$.
As we shall see, the two parameters $\mu_\infty$ and $\theta_s$ are consequently partially degenerate, with 
the degeneracy broken by the measurement of multiple acoustic peaks.

A smaller effect induced by a change in $\mu_\infty$ is a difference in baryon modulation.
As we can see in \figref{fig:CMBkspace} there is an amplitude change for modes that reach the oscillation minimum, as opposed to maximum, at recombination.
This is caused by the fact that increasing $\mu_\infty$ increases the gravitational potential and hence the baryon modulation effect.
At about the same relevance we can also see  the effect of the change in the epoch of Weyl potential decay on
the efficiency of radiation driving.
The latter two effects are difficult to see in \figref{fig:CMBlspace}  as they are sub-leading with respect to the phase shift.  They can be uncovered by cancelling the shifts due to $\mu_\infty$ and $\theta_s$ at a fiducial multipole, e.g.~the third acoustic peak.

A larger $\gamma_\infty$, on the other hand, shows three effects on the CMB power spectrum: a decrease in the amplitude of the acoustic peaks, a further decrease at scales larger than the first acoustic peak, $10\lesssim\ell\lesssim100$, and a phase shift.
First of all, the overall amplitude change comes from the driving effect. 
As we see in Eq.~(\ref{eq:Theta0-Phi}), a larger $\gamma_\infty$ gives a lower initial value of $|\Phi+\Psi|$, which decreases the value of $\Theta_0+\Psi$, hence the overall amplitude of CMB temperature fluctuation.
Notice that the integral part in Eq. \eqref{eq:Theta0-Phi} has opposite sign and approximately twice the amplitude of the initial part in GR, so the fractional difference of $C_l^{TT}$ is approximately $4$ times as the fractional difference of the initial Weyl potential Eq.~\eqref{eq:iniNewtonian} and another factor of $2$ comes from the square in the calculation of power spectrum.

On intermediate scales just larger than the first acoustic peak, the reduction in the amplitude of the Sachs-Wolfe and Doppler effects further
suppresses power in \figref{fig:CMBlspace}.  Above the horizon in the matter dominated epoch  $\Theta_0+\Psi =\Psi/3$
and so Eq.~(\ref{eq:iniNewtonian}) and the photon conservation equations predicts the rough amplitude with some reduction in the effect due to the ISW effect. 
At large scales where fluctuations are above the horizon at the transition, the SW and ISW effects from the transition add coherently and since $\mu=1$ after the transition, they cancel leaving temperature power spectrum differences that vanish as $\ell \rightarrow 0$.

Finally a deviation of $\gamma_\infty$ from unity also induces a small phase shift.   Calibrated to the third peak, the shift for $\gamma_\infty=1.2$ is $\Delta \ell \simeq 3$ and remains nearly constant throughout the acoustic peaks.
We shall see that this phase shift causes a partial degeneracy between $\mu_\infty$ and $\gamma_\infty$.

MG at late time, conversely, does not affect acoustic oscillations but rather changes the spectrum of un-lensed CMB temperature fluctuations through the ISW effect.
At the transition, the time derivative of the Weyl potential causes an enhancement of the
temperature power spectrum that is dependent on the width of the transition.   As we show in
Appendix \ref{sec:TransitionWidth}, if the transition is much sharper than $\Delta_T \sim 1$ efold, the ISW effect makes
the CMB highly sensitive to the difference between $\mu$ and $\gamma$ at early and late 
times.\footnote{Furthermore as discussed in Appendix \ref{sec:TransitionWidth},  MGCAMB  implements the switch between GR at early times and potential deviations at late times instantaneously and produces inconsistent, not merely overly sensitive, results if the parametrization is not designed to go smoothly to zero on the low redshift side of the transition (cf.~\cite{Ade:2015rim} Planck 2015 paper).}
Since such a transition is unrealistic for a model whose deviations from GR evolve on the Hubble timescale, we fix $\Delta_T=1$.

\begin{figure}
\centering
\includegraphics[width=\columnwidth]{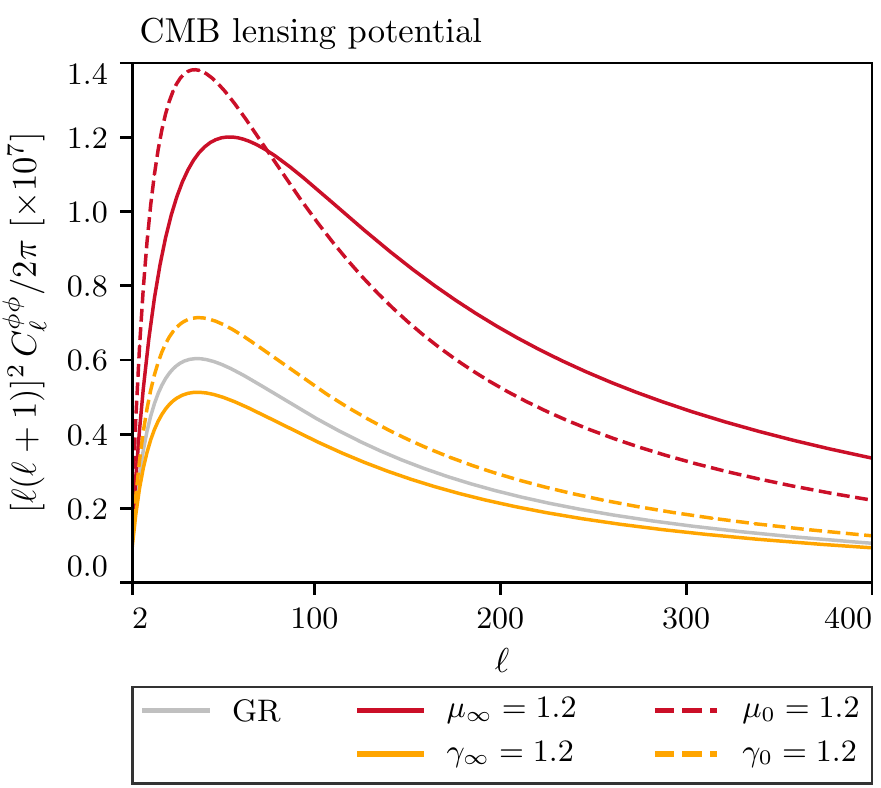}
\caption{
\label{fig:CMBlensing}
The CMB lensing potential power spectrum in the harmonic domain. Different colors correspond to different models as shown in legend.
}
\end{figure}

Both early times MG models modify the Weyl potential, as shown in \figref{fig:transfer_all}a, and thus change the lensing potential of the CMB accordingly in \figref{fig:CMBlensing}.  Raising either $\mu_\infty$ or $\mu_0$ raises the
amplitude of the lensing potential whereas $\mu_\infty$ also causes a notable change in its shape
as a result of the scale dependent enhancement of modes that entered the horizon during radiation domination as shown in \figref{fig:transfer_all}a.
Raising $\gamma_\infty$ or $\gamma_0$ decreases and increases the lensing potential respectively
with little change in the shape.  These changes to the lensing potential produce observable
effects in the smoothing of the acoustic peaks and CMB lens reconstruction.

\subsection{Weak  Lensing}
Measurements of the galaxy weak lensing (WL)  shear correlation function provide a powerful way of studying MG models.
In this section we discuss the impact of MG on weak lensing observables through the  Weyl potential.

As we discussed in the previous sections, in GR, anisotropic stress is negligible at late times so the Weyl potential power spectrum is just a re-scaling of the matter power spectrum, but this is not generally true in MG.
The difference between the two can be clearly seen comparing the two panels of \figref{fig:transfer_all}.
For this reason it is important to build lensing observables starting from the Weyl potential power spectrum.

The amplitude of the WL power spectrum is usually parametrized in terms of $\sigma_8$, the rms amplitude of linear matter density fluctuations
$\Delta_m$ convolved with a spherical tophat of radius $8\,{h}^{-1}\,{\rm Mpc}$ at $z=0$, and the matter density parameter $\Omega_m$, in their combination $S_8 \equiv \sigma_8 \Omega_m^{0.5}$.
When considering MG models we need to take into account the difference between the Weyl potential and matter density perturbations.
For this reason we replace the matter density fluctuations $\Delta_m$ with $\Delta_{\rm WL}$ where
\begin{align} \label{Eq:sigma8WL}
\Delta_{\rm WL} \equiv -\frac{k^2(\Phi+\Psi)}{8\pi G a^2 \rho_m} \,,
\end{align}
and define $\sigma_8^{\rm WL}$ using this field.
Eq. (\ref{Eq:sigma8WL}) is normalized such that, in GR and in absence of matter anisotropic stresses $\sigma_8^{\rm WL} \rightarrow \sigma_8$.
In MG it is easy to see that, in absence of matter anisotropic stresses:
\begin{align}
\sigma_8^{\rm WL} = \frac{1+\gamma}{2}\mu\,\sigma_8
\end{align}
so that its general definition extends the definition in \cite{Aghanim:2018eyx} and reduces to it when matter anisotropic stresses are negligible.

\begin{figure}
\centering
\includegraphics[width=\columnwidth]{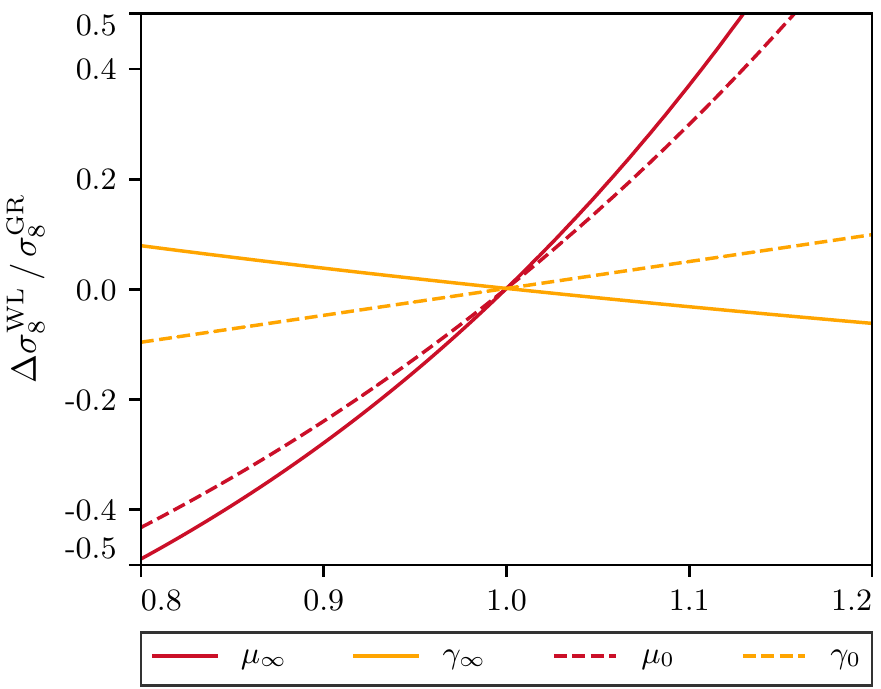}
\caption{\label{fig:Sigma8WL}
The fractional change in $\sigma_8^{\rm WL}$ from its the GR value.
Different colors correspond to different models as shown in legend.
}
\end{figure}	

We show in \figref{fig:Sigma8WL} the fractional change in $\sigma_8^{\rm WL}$ as a function of the four MG parameters around the GR value with all other parameters fixed.
As we can see this closely resembles the amplitude change of the Weyl potential at small scales, shown in \figref{fig:transfer_all}.
Raising $\mu_\infty,\mu_0,$ or $\gamma_0$ all raise the lensing observable, whereas raising $\gamma_\infty$ lowers it.
Furthermore, for the $\gamma_0$ case, the amplitude of the matter power spectrum on sub-horizon scales does not change while the amplitude of the Weyl potential tracks the change in $\gamma_0$ and illustrates why it is important to use Eq.~(\ref{Eq:sigma8WL}) as the parameter controlling WL.
Notice that the definition of $\sigma_8^{\rm WL}$ only addresses one aspect of the difference between GR and MG by incorporating the dependence on MG of the redshift zero calibration.
The amplitude of the lensing signal at a given redshift depends on the growth of perturbations and scale of the measurements.
In GR these are addressed by considering the combination $\sigma_8 \Omega_m^{0.5}$ while in MG there would be a residual dependence on $\mu$ and $\gamma$ due to the scale and redshift dependent sub horizon growth.
For compatibility with this convention we use $\sigma_8^{\rm WL} \Omega_m^{0.5}$ as a simple proxy for the WL observable in MG but test its fidelity in the $\mu_0-\gamma_0$ space directly (see \figref{fig:chi2_contour}).

\subsection{Matter power spectrum}
\begin{figure}
\centering
\includegraphics[width=\columnwidth]{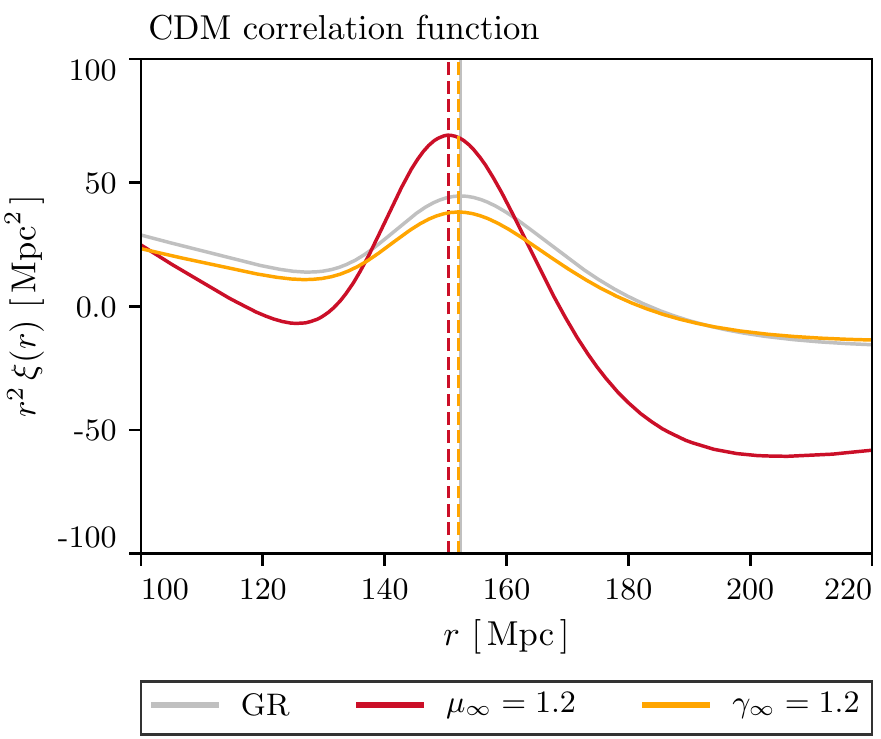}
\caption{\label{fig:BAOpeak}
The CDM spatial linear correlation function $\xi(r)$. Different colors correspond to different models as shown in legend.
Vertical lines represent the BAO peak which are slightly shifted by early time MG parameters.}
\end{figure}	
The clustering of galaxies provides another powerful and complementary probe of MG.
The modifications induced on the underlying CDM power spectrum that it traces follow closely what we discussed for the transfer functions in Sec.~\ref{sec:FullNumSolution}.
Galaxies are biased tracers of the CDM and  as shown in  \cite{Hui:2007zh,Chiang:2017vuk}, bias in the linear regime is expected to be scale dependent if the linear growth is also scale dependent as it is in MG.  
Precise modeling within parametrized approaches requires cosmological simulations  but one would expect that it qualitatively follows the response of the local growth to the linear density perturbation treated as a local background density \cite{Chiang:2017vuk}.

An observable that is more robust to these complications is the location of the BAO peak in the galaxy correlation function.
However, in case of early modified gravity, we have to take into account the fact that an acoustic phase shift induced by the decay of the gravitational potential would also shift the position of the BAO peak.

In \figref{fig:BAOpeak} we can clearly see this effect. 
We compare the CDM spatial linear correlation function in GR to the two early MG models that we consider, keeping all other cosmological parameters fixed.
The phase shift induced by $\mu_\infty$ shows as a shift in the BAO peak of the correlation function.
In our test case of $\mu_\infty=1.2$ we have that $\Delta r_{\rm peak} = 1.9 \,{\rm Mpc}$ corresponding to $\Delta r_{\rm peak}/r_{\rm peak}^{GR} = 1.2 \%$.

Since $\gamma_\infty$ also induces a phase shift we can observe a corresponding shift in the BAO peak also in the case of $\gamma_\infty=1.2$, with $\Delta r_{\rm peak} = 0.35 \,{\rm Mpc}$ and $\Delta r_{\rm peak}/r_{\rm peak}^{GR} = 0.2 \%$.  
The relative sizes of the $\gamma_\infty$ and $\mu_\infty$ shift in the BAO peak is the same as that of the phase shift in the CMB peaks.

Notice that the shape of the correlation function is also  modified in these two cases, especially in the $\mu_\infty$ case where the there is a substantial enhancement of high-$k$ power in \figref{fig:transfer_all}. 
This would provide a powerful way of testing these models once bias is understood.

We warn the reader that the measured peak position in the galaxy correlation function is usually compared to the acoustic-scale distance ratio $D_V(z)/r_{\rm drag}$, as measured by a LSS survey with effective redshift $z$.
While this holds in GR, it  is not strictly true in early MG models so that one has to check whether the difference is within the experimental error bars.

\section{Analysis Methods}\label{sec:ParamEstimation}
In this section we discuss cosmological constraints and parameter dependencies by performing MCMC parameter estimation with different datasets using a modified version of the Einstein-Boltzmann solver CAMB described in the Appendix \ref{sec:code}.	
All of the cosmology models we test have the six standard $\Lambda$CDM parameters plus the MG parameters which vary in different tests.   
The six $\Lambda$CDM parameters have standard priors and we fix the sum of neutrino masses to the minimal value~\cite{Long:2017dru}.

Since the MG parameters are introduced as phenomenological triggers for new physics in the datasets, our strategy for dataset and parameter choices is to look for datasets that are in tension with the well-measured CMB temperature power spectrum under the $\Lambda$CDM model  and to investigate the simplest MG case that might relax that tension.

Since the leading tensions under $\Lambda$CDM  involve $H_0$ and galaxy Weak Lensing (WL), we highlight these in the studies below. If MG parameters can relax these tensions, we proceed to add other cosmological datasets to see if they can provide a consistent solution.   
We also consider joint variation of MG parameters to see if together they can resolve tensions better than individually.
		
This procedure employs several datasets.
We begin with the measurements of  the high multipole  CMB temperature power spectrum from the {\it Planck} satellite~\cite{Ade:2015xua, Aghanim:2015xee} supplemented by 
the low multipole $TEB$ data which mainly constrains the optical depth $\tau$.   We refer to this baseline dataset as CMBTT.
To this we add the high multipole $EE$ and $TE$ {\it Planck} data which we call CMBpol.  
We further employ the {\it Planck} 2015 full-sky lensing potential power spectrum~\cite{Ade:2015zua} in the multipole range $40\leq \ell \leq 400$. 
At smaller angular scales CMB lensing is strongly influenced by the non-linear evolution of dark matter perturbations, we thus exclude multipoles above $\ell=400$. We refer to this dataset as CMBlens.
We indicate the dataset joining all {\it Planck} CMB datasets as CMBall = CMBTT+CMBpol+CMBlens .

The $H_0$ tension is realized by the dataset consisting of local measurements of the Hubble constant derived by the ``Supernovae, H0, for the Equation of State of dark energy'' (SH0ES) team~\cite{Riess:2016jrr} and their best estimation is $H_0=73.24\pm 1.74$ (in units of km\,s$^{-1}$\,Mpc$^{-1}$ here and throughout). We refer to this dataset as H0. 

On the other hand the WL tension leverages on the measurements of the galaxy weak lensing shear correlation function as provided by the Canada-France-Hawaii Telescope Lensing Survey (CFHTLenS)~\cite{Heymans:2013fya,Joudaki:2016mvz}. 
This dataset is referred to as WL.
We applied ultra-conservative cuts, that make CFHTLenS data insensitive to the modelling of non-linear evolution.  
We call the combination of CMBall+H0+WL = CMBtension.

Finally we include:
BAO and RSD measurements of BOSS DR12~\cite{Alam:2016hwk}, SDSS Main Galaxy Sample~\cite{Ross:2014qpa} and 6dFGS ~\cite{Beutler:2011hx};
the ``Joint Light-curve Analysis'' (JLA) Supernovae sample~\cite{Betoule:2014frx}.
We call the combination of CMBtension + BAO + SN = All.

\section{Results}\label{sec:Results}
We begin with the discussion of late times modifications to gravity, parametrized with $\mu_0$ and $\gamma_0$, and review why the preference for non-GR values appears for the CMB temperature spectrum but disappears once CMB lensing reconstruction is included.
This case has been previously considered in the literature but we correct for a problem in some of the previous implementations.  
We then consider the early times modifications to gravity, as parametrized by $\mu_\infty$ and $\gamma_\infty$, which constitute the truly new aspect of this work.   
We show that $\mu_\infty$ in particular can relax the tension with $H_0$ and WL by changing all CMB theoretical predictions in a compatible manner due to its combined effect near recombination and on lensing.  
However in our implementation where the background expansion is unmodified with respect to $\Lambda$CDM, BAO data in particular do not favor such a resolution.   
Finally we consider combinations of early and late time MG parameters to test whether non-trivial degeneracies appear and find that none of these combinations help to further reduce tensions.

\subsection{Late time modified gravity}\label{sec:LateParam}

\begin{figure}
\centering
\includegraphics[width=\columnwidth]{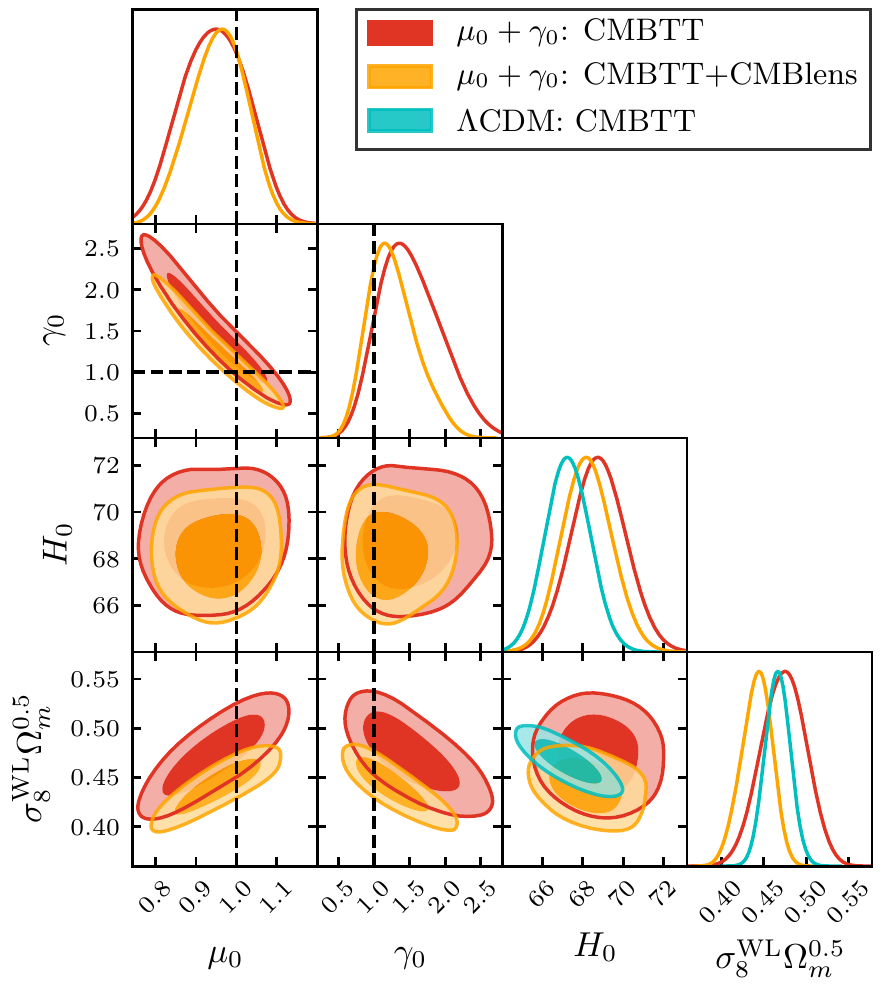}
\caption{\label{fig:show_11_MG0} 
The marginalized joint posterior of the parameters of the late time MG model, $\mu_0+\gamma_0$.
$\Lambda$CDM results are also added for comparison.
Different colors correspond to different combination of datasets, as shown in legend.
The darker and lighter shades correspond respectively to the 68\% C.L. and the 95\% C.L.
Dashed lines indicate the GR limit of the MG parameters.
}
\end{figure}

\begin{figure}
\centering
\includegraphics[width=\columnwidth]{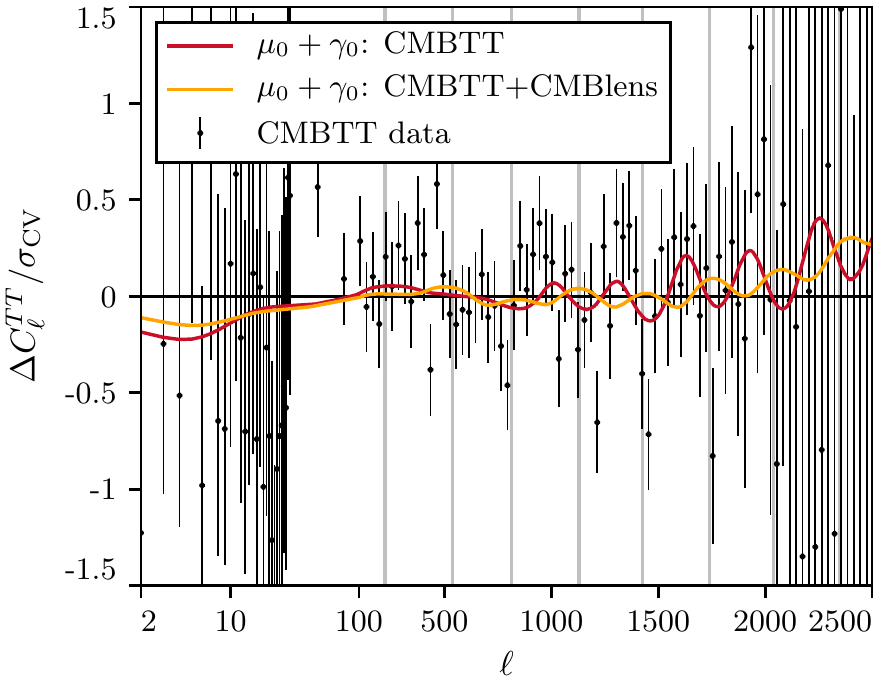}
\caption{\label{fig:TT_diff_11_MG0}
The {\it Planck} TT residuals for the best fit late time modified gravity model, relative to the best fit $\Lambda$CDM CMBTT model. 
The foreground parameters are fixed to the $\Lambda$CDM best fit values for compatibility with the foreground model removed from the data.
Different colors correspond to different combination of datasets, as shown in legend.
The residuals are normalized to $\sigma_{\rm CV}$, the cosmic variance per multipole of the best fit $\Lambda$CDM  CMBTT model (see text). 
The vertical solid lines indicate the angular position of acoustic peaks of the unlensed spectrum in the $\Lambda$CDM model.
}
\end{figure}

\begin{figure}
\centering
\includegraphics[width=\columnwidth]{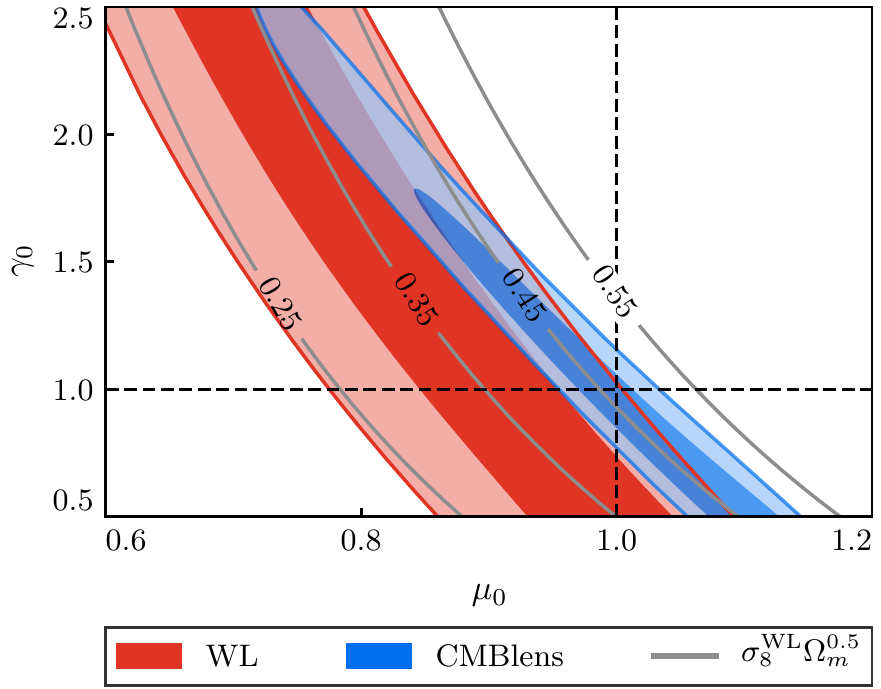}
\caption{\label{fig:chi2_contour}
The WL-only and CMBlens-only constraints on $\mu_0$ and $\gamma_0$ while keeping all other parameters fixed to their best fit $\Lambda$CDM values for CMBTT. 
The continuous lines show the iso-contours of $\sigma_8^{\rm WL}\Omega_m^{0.5}$.
The darker and lighter shades correspond respectively to the 68\% C.L. and the 95\% C.L.
Dashed lines indicate the GR limit of the MG parameters.
}
\end{figure}

We first consider the joint effects of $\mu_0$ and $\gamma_0$ since their individual effects are quite similar resulting in a strong degeneracy between the two parameters shown in \figref{fig:show_11_MG0}.
Our constraints on late time MG parameters are weaker than some of the results in the literature, especially in the degenerate direction, due to our more realistic and consistent treatment of the transition from GR at early times to these MG parameters, as discussed in Appendix \ref{sec:TransitionWidth}.
	
When the CMBTT dataset is considered alone, the late time parameters can also broaden its constraints on $H_0$ and $\sigma_8^{\rm WL}\Omega_m^{0.5}$ as shown in \figref{fig:show_11_MG0}.  
In fact a deviation from $\mu_0=\gamma_0=1$ is preferred at the 95\% C.L., as can be seen in \figref{fig:show_11_MG0}.

These results are related to coherent features in  the data residuals with respect to  the $\Lambda$CDM best fit model to the CMBTT dataset  (see \figref{fig:TT_diff_11_MG0}).    Here and below, we scale residuals to the cosmic variance per
multipole 
\begin{eqnarray}
	\sigma_{\rm CV}^{TT} &=& \sqrt{\frac{2}{2\ell+1}}C_\ell^{TT} \,, \\
	\sigma_{\rm CV}^{TE} &=& \sqrt{\frac{1}{2\ell+1}\left[C_\ell^{TT}C_\ell^{EE}+(C_\ell^{TE})^2\right]} \,,
\end{eqnarray}
with $C_\ell^{TT}$, $C_\ell^{EE}$ and $C_\ell^{TE}$ fixed to the best fit $\Lambda$CDM CMBTT model.
The data exhibit residuals that are nearly in phase with the acoustic peaks at $\ell \gtrsim 1000$ which indicate smoother acoustic peaks in the data \cite{Addison2016,Aghanim:2016sns,Obied2017}.   
This smoothness is also what is responsible for changing the inferences for a higher $H_0$ at $\ell < 1000$ and a lower $H_0$ beyond, and conversely for $\sigma_8^{\rm WL} \Omega_m^{0.5}$ given that in $\Lambda$CDM, the acoustic peak positions fix $\Omega_c h^3$ approximately.   
Furthermore the $\ell < 1000$ residuals are dominated by the low power glitch in the CMBTT data at $\ell \lesssim 30$,  shown in \figref{fig:TT_diff_11_MG0}, which explains their preference for a lower $\Omega_c h^2$ to enhance driving and a higher $H_0$.  This preference was also seen in the WMAP dataset which was comparably limited by  instrumental resolution \cite{Addison2016}.
	
It is well known that these oscillatory residuals can be better fit with a higher CMB lensing amplitude than $\Lambda$CDM implies \cite{Ade:2015xua} and this explains the preference for MG at about the same statistical significance.
In~\figref{fig:TT_diff_11_MG0}, we can see that the best fit MG parameters better accommodate the oscillatory residuals while not decreasing the agreement of the model with data below $\ell \lesssim 1000$.   
Here and below, when showing the {\it Planck} CMB best fit MG models we fix foreground parameters to their best fit 
$\Lambda$CDM values for compatibility with the data points that have that model subtracted.  On the other hand, all
constraints on MG parameters have the standard foreground parameters marginalized over.

Because of these oscillatory residuals, MG allows for a lower $\Omega_c h^2$, which also fits better the low power glitch at $\ell \lesssim 30$,  to be compatible with CMBTT and hence accommodates a higher $H_0$ and higher $\sigma_8^{\rm WL}\Omega_m^{0.5}$, where the latter reflects the fact that raising CMB lensing tends to raise WL as well.     

However it is also well known that CMBlens data do not favor raising the lensing amplitude to explain the oscillatory residuals in CMBTT \cite{Ade:2015xua} and in fact no changes in the amplitude or shape of the lens power spectrum $C_\ell^{\phi\phi}$ can reconcile them \cite{Hu:2015rva,Motloch:2018pjy}.  Consequently, once the CMB\-lens data are added the constraints on $\mu_0$ and $\gamma_0$ become compatible with the GR values and the ability to raise $H_0$ is diminished (see \figref{fig:show_11_MG0}).  
Likewise the ability to fit the oscillatory residuals in \figref{fig:TT_diff_11_MG0} also goes away.

Furthermore, the WL dataset also favor lowering not raising lensing and so would also counter the ability of
MG to fit the CMBTT oscillatory residuals.   
We can see this more directly in \figref{fig:chi2_contour} where we consider the constraints from the
CMBlens and WL datasets alone but fix the $\Lambda$CDM parameters to their best fit CMBTT values
for $\Lambda$CDM.   Note that the CMBlens and WL constrain nearly the same combination of $\mu_0$ 
and $\gamma_0$ with WL favoring somewhat lower values and hence somewhat less lensing.   
The CMBlens data are fully consistent with the GR values whereas the WL data prefer lower values,
reflecting the  tension between the CMBTT and WL datasets.   Therefore the tensions between CMBTT, CMBlens and WL data
sets cannot be resolved by raising lensing with late time MG parameters alone.   Notice also that contours of
$\sigma_8^{\rm WL} \Omega_m^{0.5}$ track the degenerate direction of the WL constraints and justify
its use as a proxy for the WL lensing observable.

\subsection{Early time modified gravity}\label{sec:EarlyParam}

\begin{figure}
\centering
\includegraphics[width=\columnwidth]{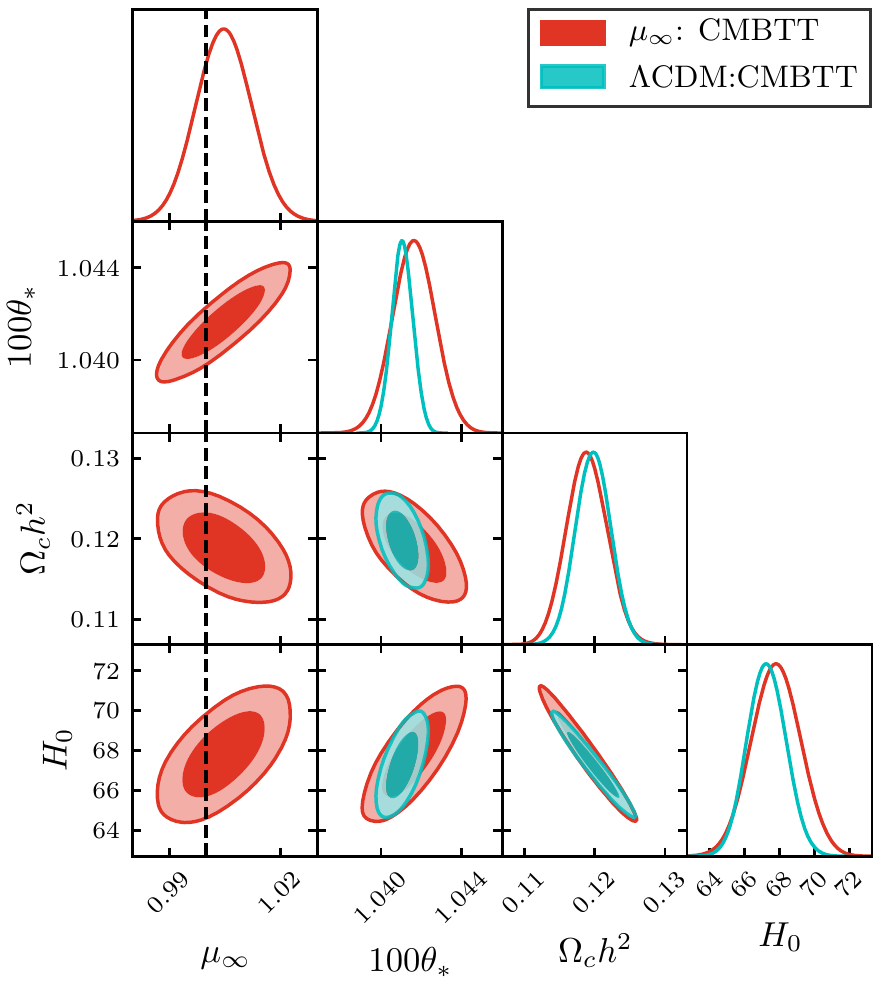}
\caption{\label{fig:show_11_muinf_param} 
The CMBTT constraints on the cosmological parameters in the $\mu_\infty$ only model.
Results of $\Lambda$CDM model are also added for comparison.
The darker and lighter shades correspond respectively to the 68\% C.L. and the 95\% C.L.
Dashed lines indicate the GR limit of the MG parameters.
}
\end{figure}	

\begin{figure}
\centering
\includegraphics[width=\columnwidth]{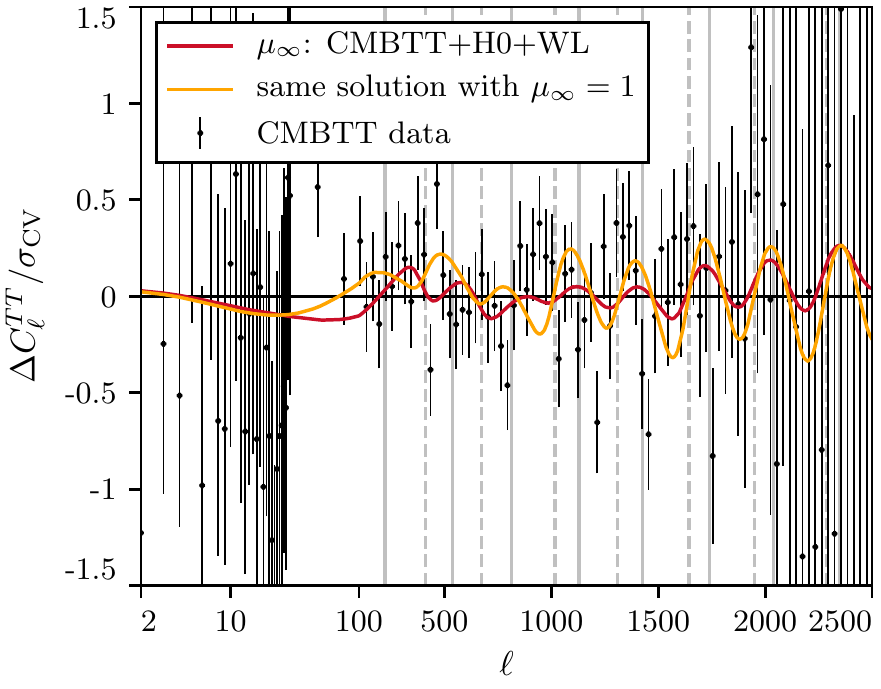}
\caption{\label{fig:TT_WL+H0+11_muinf}
TT residuals relative to the best fit $\Lambda$CDM CMBTT model, similar to \figref{fig:TT_diff_11_MG0} 
but with the best fit $\mu_\infty$-only model to CMBTT+H0+WL datasets with
$H_0\approx 70$ 
(red line).
A model with the same cosmological parameters but $\mu_\infty=1$ (orange line) shows that $\mu_\infty$ compensates
for the oscillatory fluctuations of the high $H_0$ model that are disfavored by the data.}
\end{figure}	

\begin{figure}
\centering
\includegraphics[width=\columnwidth]{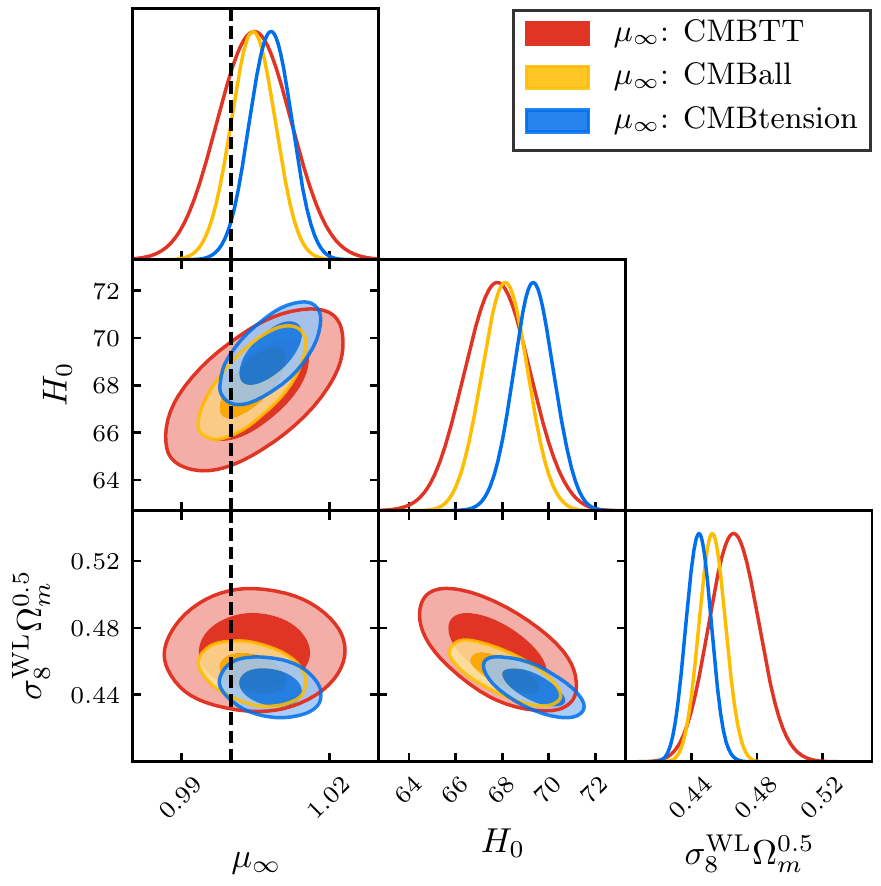}
\caption{\label{fig:show_11_muinf}
The marginalized joint posterior for the parameters in the $\mu_\infty$ only model.
Different colors correspond to different combination of datasets, as shown in legend.
The darker and lighter shades correspond respectively to the 68\% C.L. and the 95\% C.L.
The dashed lines indicate the value of $\mu_\infty$  in GR limit.  
}
\end{figure}	

\begin{figure}
\centering
\includegraphics[width=\columnwidth]{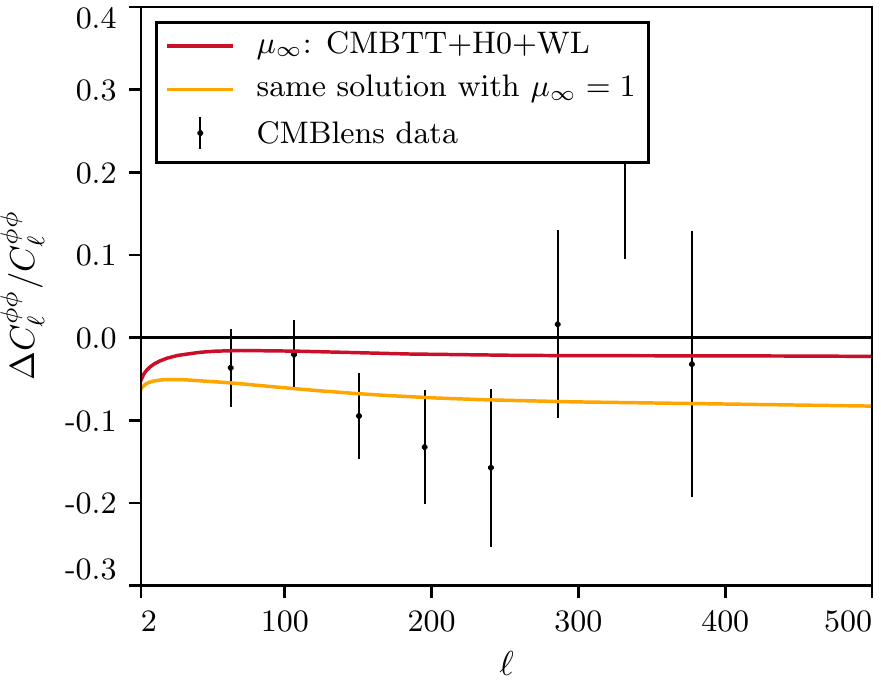}
\caption{\label{fig:phiphi_WL+H0+11_muinf}
CMB lens power spectrum residuals relative of the $\mu_\infty$ only model 
to the best fit $\Lambda$CDM CMBTT model. The models are the same as in \figref{fig:TT_WL+H0+11_muinf} as also shown in the legend.  Again $\mu_\infty$ compensates the changes due to the larger $H_0$ value.
}
\end{figure}	

\begin{figure}
\centering
\includegraphics[width=\columnwidth]{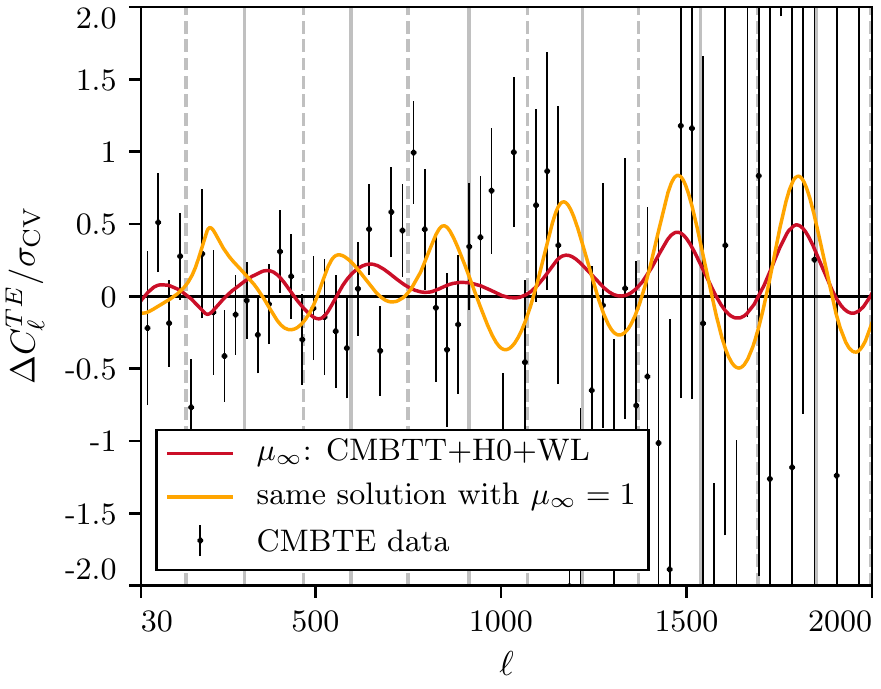}
\caption{\label{fig:TE_WL+H0+11_muinf}
TE residuals relative to the best fit $\Lambda$CDM CMBTT model, similar to \figref{fig:TT_WL+H0+11_muinf} but with TE spectrum. 
Again $\mu_\infty$ compensates the changes due to the larger $H_0$ value especially around the first minimum where the data also fluctuate low (leftmost vertical line).
}
\end{figure}	

\begin{figure}
\centering
\includegraphics[width=\columnwidth]{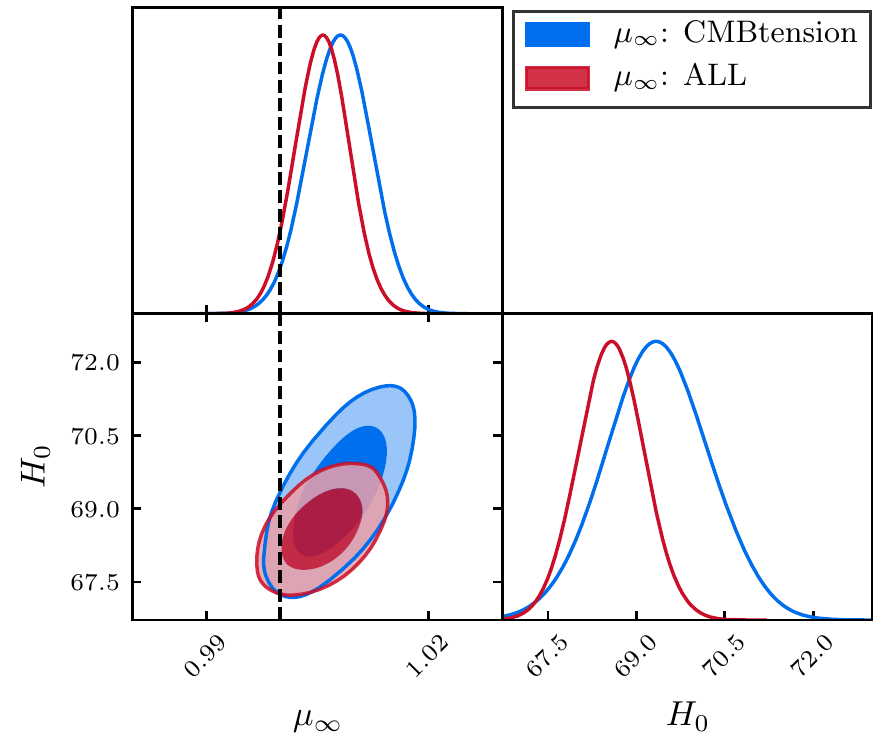}
\caption{\label{fig:show_11_muinf_ALL+-BAO}
The marginalized joint posterior for $\mu_\infty$ and $H_0$ in the $\mu_\infty$ only model.
Different colors correspond to different combination of datasets, as shown in legend.
The darker and lighter shades correspond respectively to the 68\% C.L. and the 95\% C.L.
The dashed lines indicate the value $\mu_\infty$ in GR limit.
}
\end{figure}	

\begin{figure}
\centering
\includegraphics[width=\columnwidth]{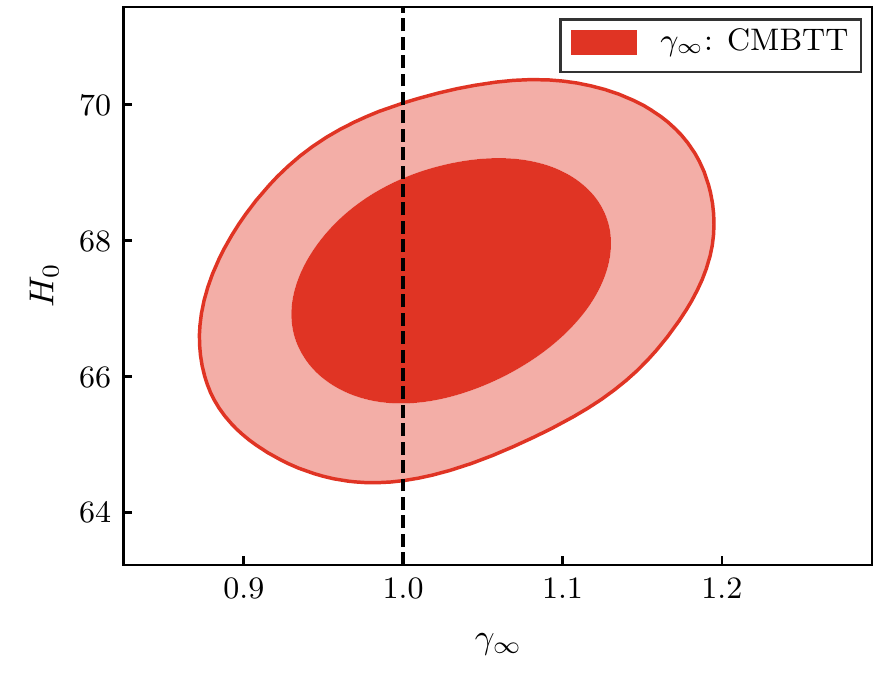}
\caption{\label{fig:show_11_gammainf_gammainf+H0}
The marginalized joint posterior for $\gamma_\infty$ and $H_0$ in the  $\gamma_\infty$ only model with the CMBTT dataset.
The darker and lighter shades correspond respectively to the 68\% C.L. and the 95\% C.L.
The dashed lines indicate the value $\gamma_\infty$ in GR limit where nearly the same range in $H_0$ is allowed.
}
\end{figure}	

The oscillatory residuals in the CMBTT data vs.\ a high $H_0$ solution cannot be explained by introducing new physics through MG to change the smoothing of the peaks by gravitational lensing since the CMBlens dataset forbids it.   
This leaves the possibility that the early time MG can change the intrinsic shape of the peaks and resolve these tensions.   
Furthermore since the study of the early time parameters $\mu_\infty$ and $\gamma_\infty$ is new to this work, we conduct a more thorough examination of their individual effects than in the previous section on late time parameters.
	
We start with $\mu_\infty$ alone and the CMBTT dataset in \figref{fig:show_11_muinf_param}.  
Here $\mu_\infty$ is strongly correlated with $\theta_*$. This is because $\mu_\infty$ leads to a phase shift in the acoustic peaks and its effect on the well-measured peak locations is partially degenerate with a change in the angular scale of the sound horizon (see Sec.~\ref{sec:CMB}).  
Since the CMBTT dataset measures multiple peaks, the degeneracy is not perfect and indeed the measurement of the peak locations provides the strongest constraint on $\mu_\infty$.  
Note also that within $\Lambda$CDM the best fit value of $\theta_*$ shifts between $\ell \lesssim 1000$ and the full dataset  to lower values by $\sim 1\sigma$ \cite{Aghanim:2016sns}.  In the $\mu_\infty$ model, this shift in peak locations can be accommodated by a change in both the angular scale and phase of the acoustic peaks.

A larger value for $\mu_\infty$ also allows a larger value of $H_0$.  In GR with the $\Lambda$CDM 
 $H_0 =67.26\pm 0.99$.  Within the bounds on $\mu_\infty$ allowed by the phase shift, $H_0=67.80 \pm 1.27$.
This comes about since $\mu_\infty$ allows another way of changing the amount of low  to high $\ell$ TT power, especially around the first few acoustic peaks, and so is partially degenerate
with radiation driving effects from $\Omega_c h^2$ and also another way of compensating the reduction in lensing due to a lower $\Omega_c h^2$.  Notice that the the correlation of the latter with $H_0$ remains largely unchanged since  the background expansion remains identical to $\Lambda$CDM and $\theta_s$ remains well constrained.   
	This has implications for BAO and SN as
we shall see, since raising $H_0$ requires the same reduction in $\Omega_m$ as in $\Lambda$CDM. Note that as in $\Lambda$CDM a higher $H_0$ solution also gives a lower $\sigma_8\Omega_m^{0.5}$ which is favored by WL data.

We can see these effects in \figref{fig:TT_WL+H0+11_muinf}.  Here we show how $\mu_\infty$ compensates for changes in the cosmological parameters of a
high $H_0 = 69.9$ best fit solution to the CMBTT+H0+WL datasets.    Notice that with the same cosmological parameters but reverting $\mu_\infty=1$,
the model exacerbates the oscillatory residuals in the data at $\ell \gtrsim 1000$, the well-known problem with
raising $H_0$ under $\Lambda$CDM. 
	
While this broadening and raising of $H_0$ values allowed by CMBTT with $\mu_\infty$ is small but significant in that it places $H_0=70$ well within the CMBTT $2\sigma$ bound, the main benefit of $\mu_\infty$ over the late time parameters is that CMB\-lens and CMB\-pol data somewhat favor rather than disfavor such a deviation from GR.
In \figref{fig:show_11_muinf}, we show the result of the CMBall combination.  Notice that the errors on $\mu_\infty$ shrink but the central
value remains the same.  After adding H0 and WL datasets, we end at $H_0 = 69.35 \pm 0.80$.
 and a preference
for $\mu_\infty>1$ at the 98.2\% C.L.  This preference comes about since the effect of raising $\mu_\infty$ and lowering $\Omega_c h^2$ nearly
compensate each other in their effect on the CMB lens power spectrum as shown in \figref{fig:phiphi_WL+H0+11_muinf} for the same model
as in  \figref{fig:TT_WL+H0+11_muinf}.
For the polarization, in $\Lambda$CDM
a low $H_0$ value is preferred in large part due to the low  TE datapoint around the first minimum ($\ell \sim 165$) \cite{Obied2017}.   This too is compensated with
$\mu_\infty$ and together with changes at higher multipole in fact brings about a better fit than $\Lambda$CDM in \figref{fig:TE_WL+H0+11_muinf}.  

On the other hand since in our parameterization the background expansion remains $\Lambda$CDM, the lower $\Omega_m$ implied by a higher 
$H_0$ causes tension with the BAO dataset.   	
After adding BAO dataset, $H_0 = 68.57 \pm 0.50$, disfavoring high values,
and the preference for $\mu_\infty>1$ is
reduced to 95.7\% C.L. (see \figref{fig:show_11_muinf_ALL+-BAO}).    In a physically motivated MG theory, we would typically expect both the background and the perturbations to be modified and so in principle this problem with 
BAO could be ameliorated by changing the expansion rate in a manner similar to adding extra
relativistic degrees of freedom under GR.
	
Next, we discuss the results with $\gamma_\infty$ as the only MG parameter.
With the CMBTT dataset, similar to  $\mu_\infty$, $\gamma_\infty$ also causes a phase shift, so it is also constrained by the locations of acoustic peaks.
However, $\gamma_\infty$ only shows a very weak correlation with $H_0$ in \figref{fig:show_11_gammainf_gammainf+H0} and cannot move $H_0$ much higher than in $\Lambda$CDM.    Its effects  at recombination share some similarity to $\mu_\infty$ in that raising it  alters the power between low and
high multipoles and accommodates a lower $\Omega_c h^2$ and higher $H_0$.   However raising $\gamma_\infty$ also lowers the CMB lens power
spectrum as shown in \figref{fig:CMBlensing} and exacerbates the oscillatory residual problem at $\ell \gtrsim 1000$.

Finally, with both $\mu_\infty$ and $\gamma_\infty$ as MG parameters, a strong correlation between $\mu_\infty$ and $\gamma_\infty$ shows up when CMBTT data is considered. This also come from the phase shift effect. 
Since both  $\mu_\infty$ and $\gamma_\infty$ shift the phase in the same direction and $\mu_\infty$ to compensate each other, they show a negative correlation with a direction consistent with the amplitudes
of their phase shifts discussed above.   Because of this anti-correlation their joint ability to ameliorate  $H_0$, CMBlens and WL
tensions is not significantly greater than that of $\mu_\infty$ alone.

\subsection{Combined early and late times MG}\label{sec:Early+Late}
\begin{figure}
\centering
\includegraphics[width=\columnwidth]{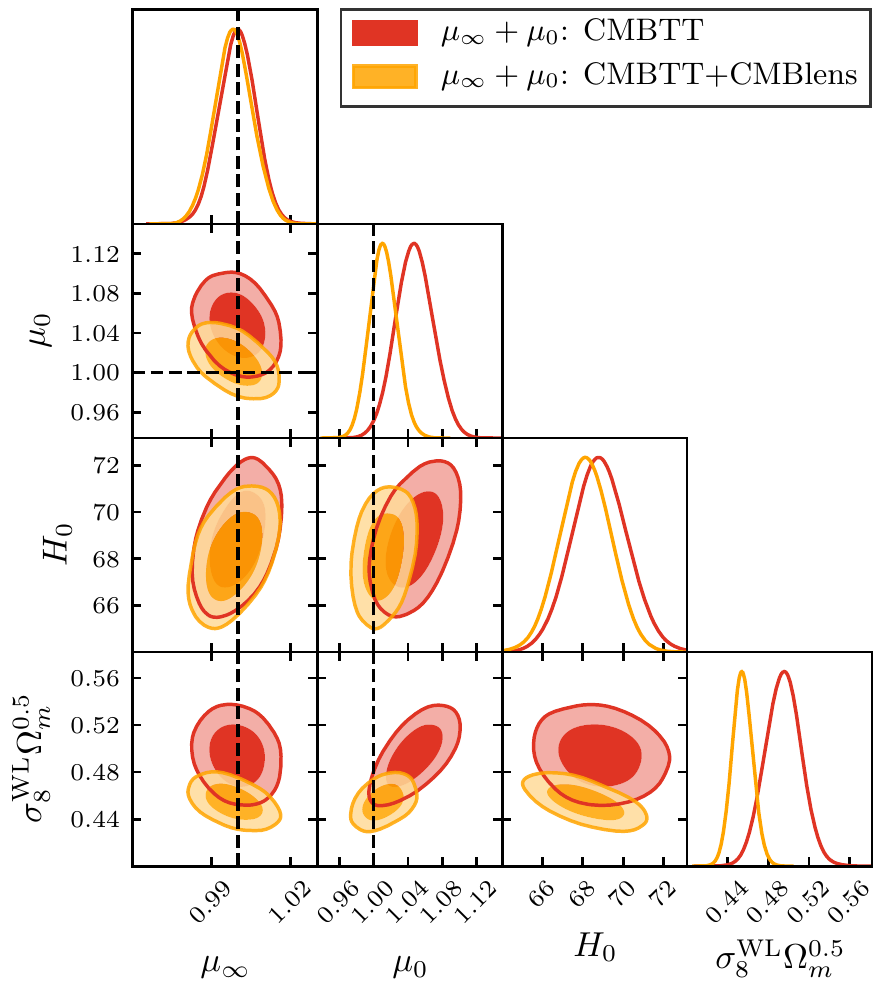}
\caption{\label{fig:show_11_mus}
The marginalized joint posterior for the parameters in the $\mu_\infty+\mu_0$ only model.
Different colors correspond to different combination of datasets, as shown in legend.
The darker and lighter shades correspond respectively to the 68\% C.L. and the 95\% C.L.
The dashed lines indicate the values of the MG parameters in the GR limit. }
\end{figure}	
\begin{figure}
\centering
\includegraphics[width=\columnwidth]{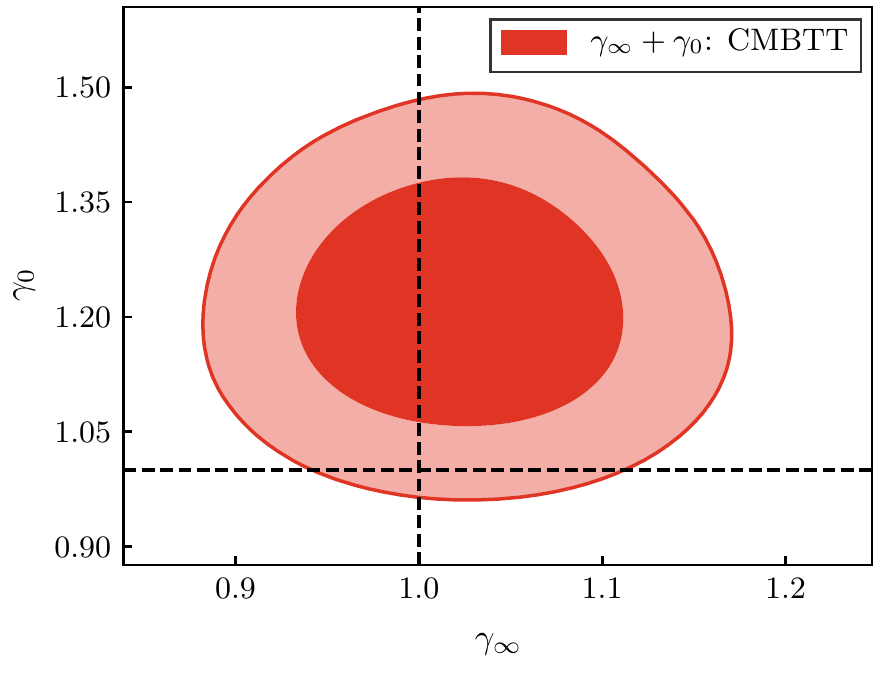}
\caption{\label{fig:show_11_gammas}
The marginalized joint posterior for $\gamma_\infty$ and $\gamma_0$ in the $\gamma_\infty+\gamma_0$ model with the CMBTT dataset.
The darker and lighter shades correspond respectively to the 68\% C.L. and the 95\% C.L. The dashed lines indicate the values of the MG parameters in the GR limit. 
}
\end{figure}	

At last we consider combinations of MG parameters together to check whether non-trivial degeneracies appear.

When we use both $\mu_\infty$ and $\mu_0$ as MG parameters, the $H_0$ tension can be further reduced when CMBTT data is considered. 
But for the same reasons that we discussed in Sec. \ref{sec:LateParam}, this resolution is also disfavored by CMBlens because of the enhancement of the lensing potential, see \figref{fig:show_11_mus}.

On the other hand, when we use both $\gamma_\infty$ and $\gamma_0$ as the MG parameters, we find that there is no correlation between them, see \figref{fig:show_11_gammas}. 
Both parameters affect the CMB temperature spectrum in independent ways and both parameters are then not favored by the data as we discussed in the previous sections.

\section{Discussion}\label{sec:Conclusions}
We study the impacts of MG on cosmological perturbation evolution and CMB power spectrum under a  phenomenological parameterization of the Poisson equations~(\ref{eq:E1}, \ref{eq:E2}).  We implement this parameterization into the Einstein-Boltzmann code CAMB.

New aspects of this work include the treatment of initial conditions in the radiation dominated epoch and the effect of MG on the CMB and matter evolution prior to recombination.   In particular, we illustrate the MG effects using step functions of time for the two MG perturbation 
parameters with an unmodified $\Lambda$CDM background to isolate their effects at early and late times.

This study is partially motivated by the existence of tensions between CMB and low redshift observables in $\Lambda$CDM.    
For the late time MG parameters $\mu_0$ and $\gamma_0$, the non-GR values are favored by CMBTT data
because the MG parameters raise the lensing potential and smooth the acoustic peaks while
also raising the Hubble constant $H_0$.  
However, this preference disappears once CMBlens data  is included because the lensing reconstruction does not favor such an enhancement.   We conclude that late time modifications alone are unlikely to resolve tensions with $\Lambda$CDM.  

Parameter tensions in $\Lambda$CDM rely heavily on the interpretation of  acoustic observables in the CMB and BAO
at recombination.   By changing the evolution of gravitational potentials, MG parameters at early times can in principle be more effective.
While changing $\gamma_\infty$ can not help to reduce tensions, changing $\mu_\infty$ can relax tensions internal to the CMB datasets and between CMBTT and $H_0$ and weak lensing.
This is achieved by changing  CMB temperature, polarization and lensing predictions in a compatible manner due to its combined effect on the acoustic oscillations and on lensing and results in a preference for $\mu_\infty> 1$ at greater than 98\% C.L. and $H_0=69.35\pm 0.80$ when combined with Hubble constant and weak lensing datasets. 
BAO data however do not favor such a resolution since the background expansion remains unchanged compared to $\Lambda$CDM in our implementation and therefore requires a lower $\Omega_m$ for a higher $H_0$. 
Combinations with other MG parameters do not further help resolve tensions.  

In a physically motivated modification of gravity, we would generally expect changes to both the background and the perturbations, leaving open the possibility that once combined this tension with BAO can be ameliorated as well.  
Moreover, our simple parameterization of MG in the perturbations with a step function in time is itself only meant to be illustrative and not a prediction based on a fundamental theory.  
Instead this study serves as a guide to the construction of physically motivated models that might resolve tensions in $\Lambda$CDM.  
We leave such studies to a future work. 

\acknowledgements
We thank 
Silvia Galli,
Macarena Lagos, 
Matteo Martinelli,
Pavel Motloch,
Samuel Passaglia,
Levon Pogosian,
Alessandra Silvestri,
Gong-Bo Zhao and
Alex Zucca
for useful comments.  MXL and WH were supported by NASA ATP NNX15AK22G and the Simons Foundation.
MR and WH were supported by U.S.\ Dept.\ of Energy contract DE-FG02-13ER41958.
This work was completed in part with resources provided by the University of Chicago Research Computing Center.

\appendix
\section{Comoving Curvature Conservation} \label{app:CoR}
In order to take the initial comoving curvature perturbation ${\cal R}$ from inflation as initial conditions for the radiation dominated universe as usual,
we need to show that it is conserved
outside the horizon 
\begin{equation}
\lim_{k/{\cal H }\rightarrow 0} \frac{{\cal R}'}{\cal R} =0
\end{equation}
 for any type of matter or modified gravity parameters $\mu$ and $\gamma$.  In this Appendix we derive the conditions under which this is true.  

Starting from the definition of the comoving curvature in terms of Newtonian gauge variables in a spatially flat universe
\begin{equation}
{\cal R} = -\Phi - \frac{\coH}{k^2} \theta,
\label{eq:Reqtheta}
\end{equation}
we use the modified gravity equations (\ref{eq:E1},\ref{eq:E2}) to obtain 
\begin{equation}
{\cal R} = 
\frac{4\pi G a^2 \mu }{k^2} \left[\gamma \Delta\rho + 3(\gamma-1)(\rho +P )\sigma \right]
- \frac{\coH}{k^2} \theta .
\end{equation}
Taking the derivative of this equation and using the  matter conservation equations (\ref{eq:matterconservation}), we obtain 
\begin{align}
{\cal R'}& = {\cal R'_{\rm GR}} + \frac{4\pi G a^2 \mu}
{ k^2 + 12 \pi G a^2 \gamma\mu (\rho+P)} \nonumber\\
&\quad \times \Big\{ C_1 \Delta\rho + C_2 (\rho+p) \frac{\theta}{\coH} + 
C_3 (\rho+P)\sigma  \nonumber\\
&\quad 
+ C_4 [(\rho+P)\sigma]'\Big\}
\end{align}
where
\begin{align}\label{eq:Ci}
C_1 &= 1-\gamma +\gamma' +\gamma\frac{\mu'}{\mu} , \nonumber\\
C_2 &= -\gamma - \frac{H H'}{4\pi G\mu(\rho+P)}, \nonumber\\
C_3 &= 3(\gamma-1)\left(1+ \frac{\mu'}{\mu}\right)  +3\gamma' ,\nonumber\\
C_4& = 3(\gamma-1),
\end{align}
Under GR all of the $C_i$ coefficients vanish leading to 
\begin{equation}
{\cal R'_{\rm GR}}= -\frac{\delta P -  ({\coH}/k^2)P' \theta - (\rho+P)\sigma }{\rho+P}.
\label{eq:RprimeGR}
\end{equation}
Note that the numerator of Eq.~(\ref{eq:RprimeGR})  is the total stress fluctuation in comoving gauge so that
under GR, ${\cal R}$ is generally conserved if stress fluctuations are negligible, {\it i.e.}~above the sound horizon.

Assuming $\mu,\gamma,\mu',\gamma' ={\cal O}(1)$ and the background only has small fluctuations from GR as well, the 
$C_i ={\cal O}(1)$ and the main difference in MG is that even without matter stress fluctuations the 
comoving curvature can evolve.   However, the extra terms in ${\cal R}'/{\cal R}$ are still suppressed by
$(k/{\coH})^2$ above the horizon as long as there are no strong cancellations in the 
 contributions to ${\cal R}$.  Note that even if the MG parameters evolve on a time scale much more rapid than
the expansion time, they just act as a superluminal sound speed with 
${\cal R}'/{\cal R}$ suppressed by $(c_{\rm eff} k/{\coH})^2$ with $c_{\rm eff}^2 = {\cal O}(\mu'/\mu,\gamma'/\gamma)$.  As $k/{\coH} \rightarrow 0$, even these cases conserve comoving
curvature.
This generalizes the proof in Ref.~\citep{Zhao:2008bn} for arbitrary forms of matter and makes explicit the connection
with the horizon scale.

\section{Boltzmann Code} \label{sec:code}
The calculations in this paper employ a modified version of the Einstein-Boltzmann solver CAMB.
Unlike the treatment in the main text, CAMB uses  synchronous gauge to represent perturbations.  
In this Appendix we detail our modifications to CAMB that mostly extend the MGCAMB implementation to early times and recombination.

\subsection{Synchronous Gauge}
The synchronous gauge of the cold dark matter, in the notation of Ref.~\citep{Ma1995}, has two spatial metric potentials, the curvature $\eta$ and the perturbation the trace of the spatial metric $h$.  These are related to the Newtonian potentials
by the gauge transformation
\begin{eqnarray}
\Psi &=& \coH (\alpha' +\alpha) , \nonumber\\
\Phi &=& \eta - \coH\alpha \,,
\end{eqnarray}
where  $\alpha = {\coH}(h' +6\eta')/2k^2$. 	
Hence, the modified Einstein equations in synchronous gauge are
\begin{align}
& \alpha'+\alpha  = -\frac{4\pi Ga^2\mu}{ k^2{\coH}}[\Delta\rho+3(\rho+P)\sigma],\label{eq:synE1}\\
& \eta - \gamma \coH {\alpha}' - (\gamma+1)\coH\alpha = \frac{12\pi Ga^2\mu}{k^2}(\rho+P)\sigma .\label{eq:synE2}
\end{align}
The combination of these two equations gives  the first equation we use to modify CAMB
\begin{equation} \label{eq:alphaeq}
\coH \alpha = \eta+\frac{4\pi Ga^2\mu}{k^2}[\gamma\Delta\rho+3(\gamma-1)(\rho+P)\sigma],
\end{equation}
to replace the Einstein equation for $h'$.
Constructed out of matter density fluctuations in synchronous gauge $\Delta  \rho = \delta \rho+ 3{\coH}(\rho+P)\theta/k^2$
and retains the same form as when constructed out of Newtonian gauge fluctuations while $\sigma$ is gauge invariant.
From this point forward in this Appendix, all matter perturbation variables are in synchronous gauge.
The synchronous matter fluctuations obey the usual conservation laws
\begin{align}
\delta \rho' + 3(\delta\rho+\delta P) &= -(\rho+P) \left( \frac{\theta}{\coH}  +\frac{h'}{2}\right), \nonumber\\
[(\rho+P)\theta]' + 4(\rho+P)\theta &= \frac{k^2}{\coH}\Big[ \delta P - (\rho+P)\sigma \Big]  \,.
\label{eq:matterconservationsynch}
\end{align}
These equations also apply to individual matter species in the absence of interactions.
We refer the reader to e.g.~Ref.~\citep{Ma1995} for the equations for baryons and photons separately in presence of Thomson scattering.
Note that the joint photon-baryon system obeys Eq.~(\ref{eq:matterconservationsynch}) which is all we require below.

For the second  equation, CAMB uses the  time-space Einstein equation.  Its modification can be derived from 
the derivative of Eq.~(\ref{eq:alphaeq}) using Eq.~(\ref{eq:synE1},{\ref{eq:matterconservationsynch})
\begin{eqnarray}\label{eq:etadot}
	{\eta}' &=& \frac{4\pi G a^2\mu}{k^2+12\pi G a^2\gamma\mu(\rho+P)}   \nonumber\\
	&&\times
		  \bigg\{ \left(1-3\frac{a^2 HH'}{k^2}\right) \gamma (\rho+P)\frac{\theta }{\coH}
- C_1 \Delta\rho  \nonumber\\
&& 
		\quad	- C_2 (\rho+P) \frac{k^2}{\coH}\alpha  - C_3 (\rho+P)\sigma \nonumber\\
			&&\quad -C_4[(\rho+P)\sigma]'
		\bigg\},
\end{eqnarray}
where the $C_i$ coefficients are defined in eq.~(\ref{eq:Ci}).
Notice that in GR only the $\theta$ source remains and that this source converts Eq.~(\ref{eq:alphaeq}) to a direct 
relation between $h'$ and $\delta\rho$ removing the velocity dependence in $\Delta\rho$.

Eq.~(\ref{eq:etadot}) corrects an error in \citep{Hojjati:2011ix}. The correction term is proportional to $(\gamma-1)\sigma$, so it  affects the results when both $\gamma\neq1$ and $\sigma$ is important. At the late times the anisotropic stress $\sigma$ is very small, so the impact of this correction is limited. However, at the early times the anisotropic stress is non-negligible, this correction has considerable influence. We also generalize the result for time-varying equations of state which is necessary for the treatment of massive neutrinos.

We also find some bugs in the publicly available Feb 2016 version of MGCAMB code. 
The bugs will affect the massive neutrino effects and the calculation of the derivatives of anisotropic stresses. 
Care should be used when employing MGCAMB in regimes where massive neutrinos or anisotropic stresses become important.
These bugs are in the process of being fixed in the public version of the MGCAMB code \cite{MGCAMBPC}.
On the other hand, in the regime where MGCAMB was mostly employed in literature, for modifications of gravity at late times, we find that such bugs have close to no effect on cosmological results.

\begin{figure*}
\centering
\includegraphics[width=\textwidth]{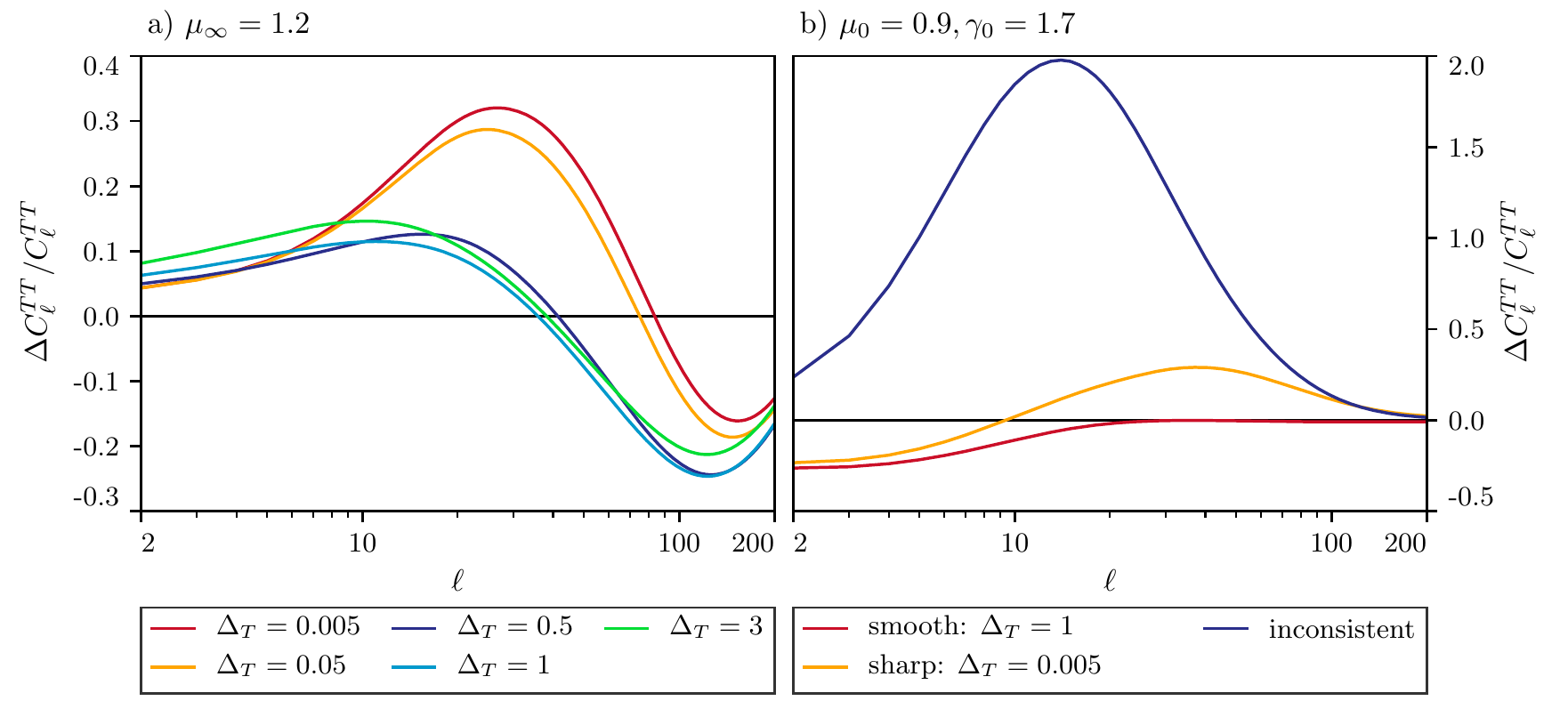}
\caption{\label{fig:width}
The comparison of the unlensed large scale CMB temperature spectrum in several MG example models with different transition widths relative to the GR spectrum. 
Left: early time MG example models ($\mu_\infty=1.2$). Right: late time MG example models along the degeneracy direction ($\mu_0=0.9$, $\gamma_0=1.7$).  Here we also show the erroneously large effect predicted from
the inconsistent implementation of an instantaneous transition employed in MGCAMB.
}
\end{figure*}

\subsection{Initial Conditions}\label{app:ini}
Here we derive the initial conditions for cosmological perturbations in synchronous gauge when the perturbations
in each $k$-mode are well outside the horizon.   In this section we
assume that $\mu$ and $\gamma$ are constant near the initial conditions.   We also assume that
the background expansion is radiation dominated with an unmodified Friedmann equation.  

Since CAMB solves
equations in conformal time $\tau$ in this section we switch from $\ln a$ to $\tau$ as the time variable.
Under the background assumption, the relationship between the two at the initial epoch is
	\begin{equation}
		\tau = \frac{2}{\omega} \left(\sqrt{1+\frac{\Omega_m}{\Omega_r}a} -1\right),
	\end{equation}
	where
		\begin{equation}
	\omega \equiv \frac{H_0\Omega_m}{\sqrt{\Omega_r}} 
	\end{equation}
	with $\rho_b+\rho_c = \Omega_m  \rho_{\rm crit}/a^3$ as the sum of the baryon and cold dark matter densities
	and $\rho_\gamma+\rho_\nu =  \Omega_r \rho_{\rm crit}/a^4$ in 	units of the present critical density $\rho_{\rm crit} = 3H_0^2/8\pi G$.
	Note that $\omega$ scales $\tau$ to its value around matter radiation equality 
	so that both $k\tau\ll1 $ and $\omega\tau\ll 1$ at the initial conditions.

Using Eq.~(\ref{eq:matterconservationsynch}) individually for the separately conserved photon-baryon, neutrino, and cold dark matter
fluids as well as  the unmodified neutrino Boltzmann equation for its anisotropic stress
\begin{eqnarray}\label{eq:synBol}
\dot{\sigma_\nu} &=& \frac{4}{15}\theta_\nu +\frac{2}{15}\dot{h} + \frac{4}{5}\dot{\eta}.
\end{eqnarray}
 and the modified 
Einstein equations, we can now solve for the initial conditions in a series expansion in $k\tau$ and $\omega\tau$.
In general we keep the terms that are sufficient to determine the next to leading order correction in
$\eta$ and $\dot h$ following the CAMB conventions.  Because $\dot h$ is derived from the modified 
equation (\ref{eq:alphaeq}) which involves $\theta$ we need to keep an extra $\omega\tau$ order in its initial condition relative to the GR result.

For adiabatic initial conditions  
\begin{equation}
\delta_\gamma=\delta_\nu=\frac{4}{3}\delta_c=\frac{4}{3}\delta_b,
\end{equation} 
we obtain
\begin{eqnarray}\label{eq:iniSyndensity}
	\frac{\eta}{\mathcal{R}} &=& -1 +\frac{15-10\mu+4\mu R_\nu}{12\mu(10\gamma+5+4\mu R_\nu)} (k\tau)^2	
	+ A_1 (\omega k^2\tau^3),
\nonumber\\
	\frac{ \delta_\gamma }{\mathcal{R}} &=& \frac{15+4\mu R_\nu}{3\mu(10\gamma+5+4\mu R_\nu)} (k\tau)^2 
	+ A_2  (\omega k^2\tau^3) , \nonumber\\
	 \frac{ \theta_\gamma}{\mathcal{R}} &=& \frac{15+4\mu R_\nu}{36\mu(10\gamma+5+4\mu R_\nu)} (k^4\tau^3)
	 + A_3   (\omega k^4\tau^4)  , \nonumber\\
	\frac{ \theta_\nu }{\mathcal{R}} &=& \frac{15+4(2+R_\nu)\mu}{36\mu(10\gamma+5+4\mu R_\nu)}  (k^4\tau^3) 
	+ A_4  (\omega k^4\tau^4)  ,\nonumber\\
	\frac{ \sigma_\nu}{\mathcal{R}}  &=& -\frac{2}{3(10\gamma+5+4\mu R_\nu)} (k\tau)^2
	+ A_5  (\omega k^2\tau^3 )
\end{eqnarray}
with
\begin{eqnarray}
A_1 &=& \frac{1}{4} A_2 + \frac{5}{4} A_5 ,\nonumber\\
A_2 &=& -\frac{30(\gamma+1)+4(13\gamma-7)\mu R_\nu+\frac{32}{15}(\mu R_\nu)^2}{\mu(45\gamma+15+8\mu R_\nu)(10\gamma+5+4\mu R_\nu)}, \nonumber\\
A_3 &=& \frac{1}{16} A_2- \frac{R_b (15 + 4 \mu R_\nu)}{48 \mu (1-R_\nu) (10\gamma+5 + 4\mu R_\nu)}  , \nonumber\\
A_4 &=& \frac{1}{16}A_2 - \frac{1}{4} A_5 ,\nonumber\\
A_5 &=& \frac{5(\gamma+1)-8\mu R_\nu}{3(45\gamma+15+8\mu R_\nu)(10\gamma+5+4\mu R_\nu)} ,
\end{eqnarray}
where $R_b=\rho_b/\rho_m$ and $R_\nu=\rho_\nu/\rho_r$ given our assumption of an unmodified expansion history.
Together with $\dot h = -2\dot{\delta_c}$, these also define the initial conditions for $\alpha$.  Note that by definition $\theta_c=0$ and by virtue of
the tight coupling between photons and baryons, $\theta_b=\theta_\gamma$ and $\sigma_\gamma=0$.  

The initial comoving curvature ${\cal R}$ from inflation coincides with $\eta$ since synchronous gauge and
comoving gauge approximately coincide outside the horizon where the density perturbations are negligible.    
Note that we have kept an $\omega k^2\tau^3$ correction to
$\eta$ for clarity and completeness even though this correction does not contribute dynamically at the desired order in either
metric fluctuation and may be omitted from the code.

\begin{figure}
\centering
\includegraphics[width=\columnwidth]{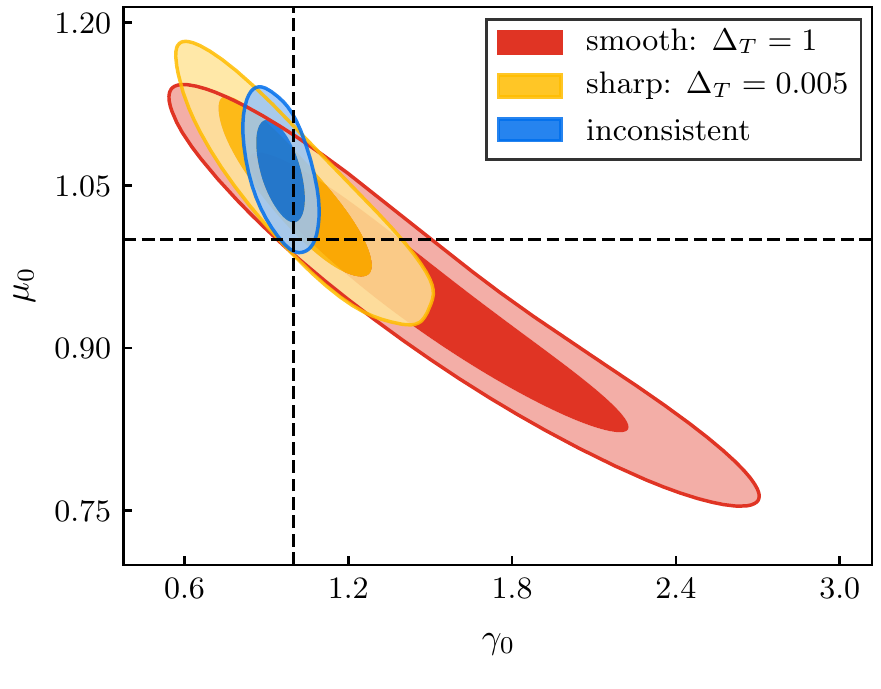}
\caption{\label{fig:show_com_MG0_mu0+gamma0}
The CMBTT constraints on late time MG parameters with different transition treatments from GR to MG.
The darker and lighter shades correspond respectively to the 68\% C.L. and the 95\% C.L.
The dashed lines indicate the values of MG parameters in GR limit.   With a sharp transition the ISW effect
shown in \figref{fig:width} breaks the degeneracy between parameters leading to unrealistically strong constraints.
The inconsistent instantaneous transition gives even stronger, but incorrect, constraints.
}
\end{figure}

\section{Impact of The Transition Width}\label{sec:TransitionWidth}

In this appendix we comment on the impact of the transition width on the CMB temperature spectrum and hence all results involving the CMBTT dataset.
For the analysis in the main text we use  $\Delta_T=1$ which we will now show leads to robust results. 

In \figref{fig:width}, we show the result of varying this width.   Around $\Delta_T=1$, there is little impact 
from varying it by factors of a few up or down.   When the transition width is much larger, $\Delta_T>3$,  
the transition will start affecting the physics of recombination and the results become highly dependent on this parameter.
When the width is very small the results are stable, but a sharp transition will introduce a relatively large
effect on the CMB power spectrum through the ISW effect, see for example the lines with $\Delta_T=0.05,\;0.005$ in \figref{fig:width}.  Since a transition that is much sharper than an efold would not be expected in a model where
modifications evolve on the Hubble time, this effect would cause the CMBTT dataset to be unrealistically
sensitive to the MG parameters.   

Furthermore results in the literature often use MGCAMB which implements an instantaneous transition at the start of the MG epoch.  
If a MG parametrization is built to have deviations from GR right after the transition time, MGCAMB will return inconsistent and possibly incorrect results.
The ISW effect depends on the time derivative of the Weyl potential and simply joining the GR and MG 
equations of motion at the transition neglects the part of its source which is the derivative of a step, 
{\it i.e.}~a delta function source. 
 In \figref{fig:width}b, we show the impact of this inconsistency on the CMB temperature power spectrum.   
 Note that unlike in \figref{fig:width}a, the inconsistent instantaneous 
 solution is not the limiting case of $\Delta_T\rightarrow 0$ but rather has a spuriously large ISW effect.   
 To use MGCAMB consistently, one must ensure that the MG functions would go back to their GR value smoothly before the transition.
 	
For example, in \cite{Ade:2015rim}, the authors used MGCAMB to implement MG without demanding
that the MG parameters  smoothly relax to the GR values.
This leads to overly tight constraints on the MG parameters.  On the other hand, the parametrization used in \cite{Aghanim:2018eyx} smoothly approaches the GR limit and by the time of the transition and leads to negligible deviations from GR so that the MGCAMB switching strategy produces consistent results.

In \figref{fig:show_com_MG0_mu0+gamma0}, we show the impact on a $\mu_0$-$\gamma_0$ MG model.   
Note that the inconsistent treatment provides much tighter constraints, especially on $\gamma_0$, than the limit of a sharp transition.    
Furthermore a consistently implemented sharp transition also provides tighter constraints than a smooth transition.    
For these reasons, we implement the smooth $\Delta_T=1$ transition for the analysis in the main text.

\clearpage
\bibliography{biblio}

\begin{thebibliography}{56}%
\makeatletter
\providecommand \@ifxundefined [1]{%
 \@ifx{#1\undefined}
}%
\providecommand \@ifnum [1]{%
 \ifnum #1\expandafter \@firstoftwo
 \else \expandafter \@secondoftwo
 \fi
}%
\providecommand \@ifx [1]{%
 \ifx #1\expandafter \@firstoftwo
 \else \expandafter \@secondoftwo
 \fi
}%
\providecommand \natexlab [1]{#1}%
\providecommand \enquote  [1]{``#1''}%
\providecommand \bibnamefont  [1]{#1}%
\providecommand \bibfnamefont [1]{#1}%
\providecommand \citenamefont [1]{#1}%
\providecommand \href@noop [0]{\@secondoftwo}%
\providecommand \href [0]{\begingroup \@sanitize@url \@href}%
\providecommand \@href[1]{\@@startlink{#1}\@@href}%
\providecommand \@@href[1]{\endgroup#1\@@endlink}%
\providecommand \@sanitize@url [0]{\catcode `\\12\catcode `\$12\catcode
  `\&12\catcode `\#12\catcode `\^12\catcode `\_12\catcode `\%12\relax}%
\providecommand \@@startlink[1]{}%
\providecommand \@@endlink[0]{}%
\providecommand \url  [0]{\begingroup\@sanitize@url \@url }%
\providecommand \@url [1]{\endgroup\@href {#1}{\urlprefix }}%
\providecommand \urlprefix  [0]{URL }%
\providecommand \Eprint [0]{\href }%
\providecommand \doibase [0]{http://dx.doi.org/}%
\providecommand \selectlanguage [0]{\@gobble}%
\providecommand \bibinfo  [0]{\@secondoftwo}%
\providecommand \bibfield  [0]{\@secondoftwo}%
\providecommand \translation [1]{[#1]}%
\providecommand \BibitemOpen [0]{}%
\providecommand \bibitemStop [0]{}%
\providecommand \bibitemNoStop [0]{.\EOS\space}%
\providecommand \EOS [0]{\spacefactor3000\relax}%
\providecommand \BibitemShut  [1]{\csname bibitem#1\endcsname}%
\let\auto@bib@innerbib\@empty
\bibitem [{\citenamefont {Riess}\ \emph {et~al.}(1998)\citenamefont {Riess}
  \emph {et~al.}}]{Riess:1998cb}%
  \BibitemOpen
  \bibfield  {author} {\bibinfo {author} {\bibfnamefont {A.~G.}\ \bibnamefont
  {Riess}} \emph {et~al.} (\bibinfo {collaboration} {Supernova Search Team}),\
  }\href {\doibase 10.1086/300499} {\bibfield  {journal} {\bibinfo  {journal}
  {Astron. J.}\ }\textbf {\bibinfo {volume} {116}},\ \bibinfo {pages} {1009}
  (\bibinfo {year} {1998})},\ \Eprint {http://arxiv.org/abs/astro-ph/9805201}
  {arXiv:astro-ph/9805201 [astro-ph]} \BibitemShut {NoStop}%
\bibitem [{\citenamefont {Perlmutter}\ \emph {et~al.}(1999)\citenamefont
  {Perlmutter} \emph {et~al.}}]{Perlmutter:1998np}%
  \BibitemOpen
  \bibfield  {author} {\bibinfo {author} {\bibfnamefont {S.}~\bibnamefont
  {Perlmutter}} \emph {et~al.} (\bibinfo {collaboration} {Supernova Cosmology
  Project}),\ }\href {\doibase 10.1086/307221} {\bibfield  {journal} {\bibinfo
  {journal} {Astrophys. J.}\ }\textbf {\bibinfo {volume} {517}},\ \bibinfo
  {pages} {565} (\bibinfo {year} {1999})},\ \Eprint
  {http://arxiv.org/abs/astro-ph/9812133} {arXiv:astro-ph/9812133 [astro-ph]}
  \BibitemShut {NoStop}%
\bibitem [{\citenamefont {Silvestri}\ and\ \citenamefont
  {Trodden}(2009)}]{Silvestri:2009hh}%
  \BibitemOpen
  \bibfield  {author} {\bibinfo {author} {\bibfnamefont {A.}~\bibnamefont
  {Silvestri}}\ and\ \bibinfo {author} {\bibfnamefont {M.}~\bibnamefont
  {Trodden}},\ }\href {\doibase 10.1088/0034-4885/72/9/096901} {\bibfield
  {journal} {\bibinfo  {journal} {Rept. Prog. Phys.}\ }\textbf {\bibinfo
  {volume} {72}},\ \bibinfo {pages} {096901} (\bibinfo {year} {2009})},\
  \Eprint {http://arxiv.org/abs/0904.0024} {arXiv:0904.0024 [astro-ph.CO]}
  \BibitemShut {NoStop}%
\bibitem [{\citenamefont {Clifton}\ \emph {et~al.}(2012)\citenamefont
  {Clifton}, \citenamefont {Ferreira}, \citenamefont {Padilla},\ and\
  \citenamefont {Skordis}}]{Clifton:2011jh}%
  \BibitemOpen
  \bibfield  {author} {\bibinfo {author} {\bibfnamefont {T.}~\bibnamefont
  {Clifton}}, \bibinfo {author} {\bibfnamefont {P.~G.}\ \bibnamefont
  {Ferreira}}, \bibinfo {author} {\bibfnamefont {A.}~\bibnamefont {Padilla}}, \
  and\ \bibinfo {author} {\bibfnamefont {C.}~\bibnamefont {Skordis}},\ }\href
  {\doibase 10.1016/j.physrep.2012.01.001} {\bibfield  {journal} {\bibinfo
  {journal} {Phys. Rept.}\ }\textbf {\bibinfo {volume} {513}},\ \bibinfo
  {pages} {1} (\bibinfo {year} {2012})},\ \Eprint
  {http://arxiv.org/abs/1106.2476} {arXiv:1106.2476 [astro-ph.CO]} \BibitemShut
  {NoStop}%
\bibitem [{\citenamefont {Joyce}\ \emph {et~al.}(2015)\citenamefont {Joyce},
  \citenamefont {Jain}, \citenamefont {Khoury},\ and\ \citenamefont
  {Trodden}}]{Joyce:2014kja}%
  \BibitemOpen
  \bibfield  {author} {\bibinfo {author} {\bibfnamefont {A.}~\bibnamefont
  {Joyce}}, \bibinfo {author} {\bibfnamefont {B.}~\bibnamefont {Jain}},
  \bibinfo {author} {\bibfnamefont {J.}~\bibnamefont {Khoury}}, \ and\ \bibinfo
  {author} {\bibfnamefont {M.}~\bibnamefont {Trodden}},\ }\href {\doibase
  10.1016/j.physrep.2014.12.002} {\bibfield  {journal} {\bibinfo  {journal}
  {Phys. Rept.}\ }\textbf {\bibinfo {volume} {568}},\ \bibinfo {pages} {1}
  (\bibinfo {year} {2015})},\ \Eprint {http://arxiv.org/abs/1407.0059}
  {arXiv:1407.0059 [astro-ph.CO]} \BibitemShut {NoStop}%
\bibitem [{\citenamefont {Ade}\ \emph {et~al.}(2016{\natexlab{a}})\citenamefont
  {Ade} \emph {et~al.}}]{Ade:2015rim}%
  \BibitemOpen
  \bibfield  {author} {\bibinfo {author} {\bibfnamefont {P.~A.~R.}\
  \bibnamefont {Ade}} \emph {et~al.} (\bibinfo {collaboration} {Planck}),\
  }\href {\doibase 10.1051/0004-6361/201525814} {\bibfield  {journal} {\bibinfo
   {journal} {Astron. Astrophys.}\ }\textbf {\bibinfo {volume} {594}},\
  \bibinfo {pages} {A14} (\bibinfo {year} {2016}{\natexlab{a}})},\ \Eprint
  {http://arxiv.org/abs/1502.01590} {arXiv:1502.01590 [astro-ph.CO]}
  \BibitemShut {NoStop}%
\bibitem [{\citenamefont {Joudaki}\ \emph
  {et~al.}(2017{\natexlab{a}})\citenamefont {Joudaki} \emph
  {et~al.}}]{Joudaki:2016kym}%
  \BibitemOpen
  \bibfield  {author} {\bibinfo {author} {\bibfnamefont {S.}~\bibnamefont
  {Joudaki}} \emph {et~al.},\ }\href {\doibase 10.1093/mnras/stx998} {\bibfield
   {journal} {\bibinfo  {journal} {Mon. Not. Roy. Astron. Soc.}\ }\textbf
  {\bibinfo {volume} {471}},\ \bibinfo {pages} {1259} (\bibinfo {year}
  {2017}{\natexlab{a}})},\ \Eprint {http://arxiv.org/abs/1610.04606}
  {arXiv:1610.04606 [astro-ph.CO]} \BibitemShut {NoStop}%
\bibitem [{\citenamefont {Aghanim}\ \emph {et~al.}(2018)\citenamefont {Aghanim}
  \emph {et~al.}}]{Aghanim:2018eyx}%
  \BibitemOpen
  \bibfield  {author} {\bibinfo {author} {\bibfnamefont {N.}~\bibnamefont
  {Aghanim}} \emph {et~al.} (\bibinfo {collaboration} {Planck}),\ }\href@noop
  {} {\  (\bibinfo {year} {2018})},\ \Eprint {http://arxiv.org/abs/1807.06209}
  {arXiv:1807.06209 [astro-ph.CO]} \BibitemShut {NoStop}%
\bibitem [{\citenamefont {Alam}\ \emph {et~al.}(2017)\citenamefont {Alam} \emph
  {et~al.}}]{Alam:2016hwk}%
  \BibitemOpen
  \bibfield  {author} {\bibinfo {author} {\bibfnamefont {S.}~\bibnamefont
  {Alam}} \emph {et~al.} (\bibinfo {collaboration} {BOSS}),\ }\href {\doibase
  10.1093/mnras/stx721} {\bibfield  {journal} {\bibinfo  {journal} {Mon. Not.
  Roy. Astron. Soc.}\ }\textbf {\bibinfo {volume} {470}},\ \bibinfo {pages}
  {2617} (\bibinfo {year} {2017})},\ \Eprint {http://arxiv.org/abs/1607.03155}
  {arXiv:1607.03155 [astro-ph.CO]} \BibitemShut {NoStop}%
\bibitem [{\citenamefont {Abazajian}\ \emph {et~al.}(2016)\citenamefont
  {Abazajian} \emph {et~al.}}]{Abazajian:2016yjj}%
  \BibitemOpen
  \bibfield  {author} {\bibinfo {author} {\bibfnamefont {K.~N.}\ \bibnamefont
  {Abazajian}} \emph {et~al.} (\bibinfo {collaboration} {CMB-S4}),\ }\href@noop
  {} {\  (\bibinfo {year} {2016})},\ \Eprint {http://arxiv.org/abs/1610.02743}
  {arXiv:1610.02743 [astro-ph.CO]} \BibitemShut {NoStop}%
\bibitem [{\citenamefont {Laureijs}\ \emph {et~al.}(2011)\citenamefont
  {Laureijs} \emph {et~al.}}]{Laureijs:2011gra}%
  \BibitemOpen
  \bibfield  {author} {\bibinfo {author} {\bibfnamefont {R.}~\bibnamefont
  {Laureijs}} \emph {et~al.} (\bibinfo {collaboration} {EUCLID}),\ }\href@noop
  {} {\  (\bibinfo {year} {2011})},\ \Eprint {http://arxiv.org/abs/1110.3193}
  {arXiv:1110.3193 [astro-ph.CO]} \BibitemShut {NoStop}%
\bibitem [{\citenamefont {Abell}\ \emph {et~al.}(2009)\citenamefont {Abell}
  \emph {et~al.}}]{Abell:2009aa}%
  \BibitemOpen
  \bibfield  {author} {\bibinfo {author} {\bibfnamefont {P.~A.}\ \bibnamefont
  {Abell}} \emph {et~al.} (\bibinfo {collaboration} {LSST Science, LSST
  Project}),\ }\href@noop {} {\  (\bibinfo {year} {2009})},\ \Eprint
  {http://arxiv.org/abs/0912.0201} {arXiv:0912.0201 [astro-ph.IM]} \BibitemShut
  {NoStop}%
\bibitem [{\citenamefont {Zhao}\ \emph
  {et~al.}(2009{\natexlab{a}})\citenamefont {Zhao}, \citenamefont {Pogosian},
  \citenamefont {Silvestri},\ and\ \citenamefont {Zylberberg}}]{Zhao:2008bn}%
  \BibitemOpen
  \bibfield  {author} {\bibinfo {author} {\bibfnamefont {G.-B.}\ \bibnamefont
  {Zhao}}, \bibinfo {author} {\bibfnamefont {L.}~\bibnamefont {Pogosian}},
  \bibinfo {author} {\bibfnamefont {A.}~\bibnamefont {Silvestri}}, \ and\
  \bibinfo {author} {\bibfnamefont {J.}~\bibnamefont {Zylberberg}},\ }\href
  {\doibase 10.1103/PhysRevD.79.083513} {\bibfield  {journal} {\bibinfo
  {journal} {Phys. Rev.}\ }\textbf {\bibinfo {volume} {D79}},\ \bibinfo {pages}
  {083513} (\bibinfo {year} {2009}{\natexlab{a}})},\ \Eprint
  {http://arxiv.org/abs/0809.3791} {arXiv:0809.3791 [astro-ph]} \BibitemShut
  {NoStop}%
\bibitem [{\citenamefont {Zhao}\ \emph
  {et~al.}(2009{\natexlab{b}})\citenamefont {Zhao}, \citenamefont {Pogosian},
  \citenamefont {Silvestri},\ and\ \citenamefont {Zylberberg}}]{Zhao:2009fn}%
  \BibitemOpen
  \bibfield  {author} {\bibinfo {author} {\bibfnamefont {G.-B.}\ \bibnamefont
  {Zhao}}, \bibinfo {author} {\bibfnamefont {L.}~\bibnamefont {Pogosian}},
  \bibinfo {author} {\bibfnamefont {A.}~\bibnamefont {Silvestri}}, \ and\
  \bibinfo {author} {\bibfnamefont {J.}~\bibnamefont {Zylberberg}},\ }\href
  {\doibase 10.1103/PhysRevLett.103.241301} {\bibfield  {journal} {\bibinfo
  {journal} {Phys. Rev. Lett.}\ }\textbf {\bibinfo {volume} {103}},\ \bibinfo
  {pages} {241301} (\bibinfo {year} {2009}{\natexlab{b}})},\ \Eprint
  {http://arxiv.org/abs/0905.1326} {arXiv:0905.1326 [astro-ph.CO]} \BibitemShut
  {NoStop}%
\bibitem [{\citenamefont {Hojjati}\ \emph {et~al.}(2012)\citenamefont
  {Hojjati}, \citenamefont {Zhao}, \citenamefont {Pogosian}, \citenamefont
  {Silvestri}, \citenamefont {Crittenden},\ and\ \citenamefont
  {Koyama}}]{Hojjati:2011xd}%
  \BibitemOpen
  \bibfield  {author} {\bibinfo {author} {\bibfnamefont {A.}~\bibnamefont
  {Hojjati}}, \bibinfo {author} {\bibfnamefont {G.-B.}\ \bibnamefont {Zhao}},
  \bibinfo {author} {\bibfnamefont {L.}~\bibnamefont {Pogosian}}, \bibinfo
  {author} {\bibfnamefont {A.}~\bibnamefont {Silvestri}}, \bibinfo {author}
  {\bibfnamefont {R.}~\bibnamefont {Crittenden}}, \ and\ \bibinfo {author}
  {\bibfnamefont {K.}~\bibnamefont {Koyama}},\ }\href {\doibase
  10.1103/PhysRevD.85.043508} {\bibfield  {journal} {\bibinfo  {journal} {Phys.
  Rev.}\ }\textbf {\bibinfo {volume} {D85}},\ \bibinfo {pages} {043508}
  (\bibinfo {year} {2012})},\ \Eprint {http://arxiv.org/abs/1111.3960}
  {arXiv:1111.3960 [astro-ph.CO]} \BibitemShut {NoStop}%
\bibitem [{\citenamefont {Asaba}\ \emph {et~al.}(2013)\citenamefont {Asaba},
  \citenamefont {Hikage}, \citenamefont {Koyama}, \citenamefont {Zhao},
  \citenamefont {Hojjati},\ and\ \citenamefont {Pogosian}}]{Asaba:2013mxj}%
  \BibitemOpen
  \bibfield  {author} {\bibinfo {author} {\bibfnamefont {S.}~\bibnamefont
  {Asaba}}, \bibinfo {author} {\bibfnamefont {C.}~\bibnamefont {Hikage}},
  \bibinfo {author} {\bibfnamefont {K.}~\bibnamefont {Koyama}}, \bibinfo
  {author} {\bibfnamefont {G.-B.}\ \bibnamefont {Zhao}}, \bibinfo {author}
  {\bibfnamefont {A.}~\bibnamefont {Hojjati}}, \ and\ \bibinfo {author}
  {\bibfnamefont {L.}~\bibnamefont {Pogosian}},\ }\href {\doibase
  10.1088/1475-7516/2013/08/029} {\bibfield  {journal} {\bibinfo  {journal}
  {JCAP}\ }\textbf {\bibinfo {volume} {1308}},\ \bibinfo {pages} {029}
  (\bibinfo {year} {2013})},\ \Eprint {http://arxiv.org/abs/1306.2546}
  {arXiv:1306.2546 [astro-ph.CO]} \BibitemShut {NoStop}%
\bibitem [{\citenamefont {Caldwell}\ \emph {et~al.}(2007)\citenamefont
  {Caldwell}, \citenamefont {Cooray},\ and\ \citenamefont
  {Melchiorri}}]{Caldwell:2007cw}%
  \BibitemOpen
  \bibfield  {author} {\bibinfo {author} {\bibfnamefont {R.}~\bibnamefont
  {Caldwell}}, \bibinfo {author} {\bibfnamefont {A.}~\bibnamefont {Cooray}}, \
  and\ \bibinfo {author} {\bibfnamefont {A.}~\bibnamefont {Melchiorri}},\
  }\href {\doibase 10.1103/PhysRevD.76.023507} {\bibfield  {journal} {\bibinfo
  {journal} {Phys. Rev.}\ }\textbf {\bibinfo {volume} {D76}},\ \bibinfo {pages}
  {023507} (\bibinfo {year} {2007})},\ \Eprint
  {http://arxiv.org/abs/astro-ph/0703375} {arXiv:astro-ph/0703375 [ASTRO-PH]}
  \BibitemShut {NoStop}%
\bibitem [{\citenamefont {Amendola}\ \emph {et~al.}(2008)\citenamefont
  {Amendola}, \citenamefont {Kunz},\ and\ \citenamefont
  {Sapone}}]{Amendola:2007rr}%
  \BibitemOpen
  \bibfield  {author} {\bibinfo {author} {\bibfnamefont {L.}~\bibnamefont
  {Amendola}}, \bibinfo {author} {\bibfnamefont {M.}~\bibnamefont {Kunz}}, \
  and\ \bibinfo {author} {\bibfnamefont {D.}~\bibnamefont {Sapone}},\ }\href
  {\doibase 10.1088/1475-7516/2008/04/013} {\bibfield  {journal} {\bibinfo
  {journal} {JCAP}\ }\textbf {\bibinfo {volume} {0804}},\ \bibinfo {pages}
  {013} (\bibinfo {year} {2008})},\ \Eprint {http://arxiv.org/abs/0704.2421}
  {arXiv:0704.2421 [astro-ph]} \BibitemShut {NoStop}%
\bibitem [{\citenamefont {Hu}\ and\ \citenamefont {Sawicki}(2007)}]{Hu:2007pj}%
  \BibitemOpen
  \bibfield  {author} {\bibinfo {author} {\bibfnamefont {W.}~\bibnamefont
  {Hu}}\ and\ \bibinfo {author} {\bibfnamefont {I.}~\bibnamefont {Sawicki}},\
  }\href {\doibase 10.1103/PhysRevD.76.104043} {\bibfield  {journal} {\bibinfo
  {journal} {Phys. Rev.}\ }\textbf {\bibinfo {volume} {D76}},\ \bibinfo {pages}
  {104043} (\bibinfo {year} {2007})},\ \Eprint {http://arxiv.org/abs/0708.1190}
  {arXiv:0708.1190 [astro-ph]} \BibitemShut {NoStop}%
\bibitem [{\citenamefont {Bertschinger}\ and\ \citenamefont
  {Zukin}(2008)}]{Bertschinger:2008zb}%
  \BibitemOpen
  \bibfield  {author} {\bibinfo {author} {\bibfnamefont {E.}~\bibnamefont
  {Bertschinger}}\ and\ \bibinfo {author} {\bibfnamefont {P.}~\bibnamefont
  {Zukin}},\ }\href {\doibase 10.1103/PhysRevD.78.024015} {\bibfield  {journal}
  {\bibinfo  {journal} {Phys. Rev.}\ }\textbf {\bibinfo {volume} {D78}},\
  \bibinfo {pages} {024015} (\bibinfo {year} {2008})},\ \Eprint
  {http://arxiv.org/abs/0801.2431} {arXiv:0801.2431 [astro-ph]} \BibitemShut
  {NoStop}%
\bibitem [{\citenamefont {Pogosian}\ \emph {et~al.}(2010)\citenamefont
  {Pogosian}, \citenamefont {Silvestri}, \citenamefont {Koyama},\ and\
  \citenamefont {Zhao}}]{Pogosian:2010tj}%
  \BibitemOpen
  \bibfield  {author} {\bibinfo {author} {\bibfnamefont {L.}~\bibnamefont
  {Pogosian}}, \bibinfo {author} {\bibfnamefont {A.}~\bibnamefont {Silvestri}},
  \bibinfo {author} {\bibfnamefont {K.}~\bibnamefont {Koyama}}, \ and\ \bibinfo
  {author} {\bibfnamefont {G.-B.}\ \bibnamefont {Zhao}},\ }\href {\doibase
  10.1103/PhysRevD.81.104023} {\bibfield  {journal} {\bibinfo  {journal} {Phys.
  Rev.}\ }\textbf {\bibinfo {volume} {D81}},\ \bibinfo {pages} {104023}
  (\bibinfo {year} {2010})},\ \Eprint {http://arxiv.org/abs/1002.2382}
  {arXiv:1002.2382 [astro-ph.CO]} \BibitemShut {NoStop}%
\bibitem [{\citenamefont {Pogosian}\ and\ \citenamefont
  {Silvestri}(2016)}]{Pogosian:2016pwr}%
  \BibitemOpen
  \bibfield  {author} {\bibinfo {author} {\bibfnamefont {L.}~\bibnamefont
  {Pogosian}}\ and\ \bibinfo {author} {\bibfnamefont {A.}~\bibnamefont
  {Silvestri}},\ }\href {\doibase 10.1103/PhysRevD.94.104014} {\bibfield
  {journal} {\bibinfo  {journal} {Phys. Rev.}\ }\textbf {\bibinfo {volume}
  {D94}},\ \bibinfo {pages} {104014} (\bibinfo {year} {2016})},\ \Eprint
  {http://arxiv.org/abs/1606.05339} {arXiv:1606.05339 [astro-ph.CO]}
  \BibitemShut {NoStop}%
\bibitem [{\citenamefont {Hojjati}\ \emph {et~al.}(2016)\citenamefont
  {Hojjati}, \citenamefont {Plahn}, \citenamefont {Zucca}, \citenamefont
  {Pogosian}, \citenamefont {Brax}, \citenamefont {Davis},\ and\ \citenamefont
  {Zhao}}]{Hojjati:2015ojt}%
  \BibitemOpen
  \bibfield  {author} {\bibinfo {author} {\bibfnamefont {A.}~\bibnamefont
  {Hojjati}}, \bibinfo {author} {\bibfnamefont {A.}~\bibnamefont {Plahn}},
  \bibinfo {author} {\bibfnamefont {A.}~\bibnamefont {Zucca}}, \bibinfo
  {author} {\bibfnamefont {L.}~\bibnamefont {Pogosian}}, \bibinfo {author}
  {\bibfnamefont {P.}~\bibnamefont {Brax}}, \bibinfo {author} {\bibfnamefont
  {A.-C.}\ \bibnamefont {Davis}}, \ and\ \bibinfo {author} {\bibfnamefont
  {G.-B.}\ \bibnamefont {Zhao}},\ }\href {\doibase 10.1103/PhysRevD.93.043531}
  {\bibfield  {journal} {\bibinfo  {journal} {Phys. Rev.}\ }\textbf {\bibinfo
  {volume} {D93}},\ \bibinfo {pages} {043531} (\bibinfo {year} {2016})},\
  \Eprint {http://arxiv.org/abs/1511.05962} {arXiv:1511.05962 [astro-ph.CO]}
  \BibitemShut {NoStop}%
\bibitem [{\citenamefont {Zhao}\ \emph {et~al.}(2017)\citenamefont {Zhao} \emph
  {et~al.}}]{Zhao:2017cud}%
  \BibitemOpen
  \bibfield  {author} {\bibinfo {author} {\bibfnamefont {G.-B.}\ \bibnamefont
  {Zhao}} \emph {et~al.},\ }\href {\doibase 10.1038/s41550-017-0216-z}
  {\bibfield  {journal} {\bibinfo  {journal} {Nat. Astron.}\ }\textbf {\bibinfo
  {volume} {1}},\ \bibinfo {pages} {627} (\bibinfo {year} {2017})},\ \Eprint
  {http://arxiv.org/abs/1701.08165} {arXiv:1701.08165 [astro-ph.CO]}
  \BibitemShut {NoStop}%
\bibitem [{\citenamefont {Casas}\ \emph {et~al.}(2017)\citenamefont {Casas},
  \citenamefont {Kunz}, \citenamefont {Martinelli},\ and\ \citenamefont
  {Pettorino}}]{Casas:2017eob}%
  \BibitemOpen
  \bibfield  {author} {\bibinfo {author} {\bibfnamefont {S.}~\bibnamefont
  {Casas}}, \bibinfo {author} {\bibfnamefont {M.}~\bibnamefont {Kunz}},
  \bibinfo {author} {\bibfnamefont {M.}~\bibnamefont {Martinelli}}, \ and\
  \bibinfo {author} {\bibfnamefont {V.}~\bibnamefont {Pettorino}},\ }\href
  {\doibase 10.1016/j.dark.2017.09.009} {\bibfield  {journal} {\bibinfo
  {journal} {Phys. Dark Univ.}\ }\textbf {\bibinfo {volume} {18}},\ \bibinfo
  {pages} {73} (\bibinfo {year} {2017})},\ \Eprint
  {http://arxiv.org/abs/1703.01271} {arXiv:1703.01271 [astro-ph.CO]}
  \BibitemShut {NoStop}%
\bibitem [{\citenamefont {Peirone}\ \emph {et~al.}(2018)\citenamefont
  {Peirone}, \citenamefont {Koyama}, \citenamefont {Pogosian}, \citenamefont
  {Raveri},\ and\ \citenamefont {Silvestri}}]{Peirone:2017ywi}%
  \BibitemOpen
  \bibfield  {author} {\bibinfo {author} {\bibfnamefont {S.}~\bibnamefont
  {Peirone}}, \bibinfo {author} {\bibfnamefont {K.}~\bibnamefont {Koyama}},
  \bibinfo {author} {\bibfnamefont {L.}~\bibnamefont {Pogosian}}, \bibinfo
  {author} {\bibfnamefont {M.}~\bibnamefont {Raveri}}, \ and\ \bibinfo {author}
  {\bibfnamefont {A.}~\bibnamefont {Silvestri}},\ }\href {\doibase
  10.1103/PhysRevD.97.043519} {\bibfield  {journal} {\bibinfo  {journal} {Phys.
  Rev.}\ }\textbf {\bibinfo {volume} {D97}},\ \bibinfo {pages} {043519}
  (\bibinfo {year} {2018})},\ \Eprint {http://arxiv.org/abs/1712.00444}
  {arXiv:1712.00444 [astro-ph.CO]} \BibitemShut {NoStop}%
\bibitem [{\citenamefont {Espejo}\ \emph {et~al.}(2018)\citenamefont {Espejo},
  \citenamefont {Peirone}, \citenamefont {Raveri}, \citenamefont {Koyama},
  \citenamefont {Pogosian},\ and\ \citenamefont {Silvestri}}]{Espejo:2018hxa}%
  \BibitemOpen
  \bibfield  {author} {\bibinfo {author} {\bibfnamefont {J.}~\bibnamefont
  {Espejo}}, \bibinfo {author} {\bibfnamefont {S.}~\bibnamefont {Peirone}},
  \bibinfo {author} {\bibfnamefont {M.}~\bibnamefont {Raveri}}, \bibinfo
  {author} {\bibfnamefont {K.}~\bibnamefont {Koyama}}, \bibinfo {author}
  {\bibfnamefont {L.}~\bibnamefont {Pogosian}}, \ and\ \bibinfo {author}
  {\bibfnamefont {A.}~\bibnamefont {Silvestri}},\ }\href@noop {} {\  (\bibinfo
  {year} {2018})},\ \Eprint {http://arxiv.org/abs/1809.01121} {arXiv:1809.01121
  [astro-ph.CO]} \BibitemShut {NoStop}%
\bibitem [{\citenamefont {{Lewis}}\ \emph {et~al.}(2000)\citenamefont
  {{Lewis}}, \citenamefont {{Challinor}},\ and\ \citenamefont
  {{Lasenby}}}]{Lewis2000}%
  \BibitemOpen
  \bibfield  {author} {\bibinfo {author} {\bibfnamefont {A.}~\bibnamefont
  {{Lewis}}}, \bibinfo {author} {\bibfnamefont {A.}~\bibnamefont
  {{Challinor}}}, \ and\ \bibinfo {author} {\bibfnamefont {A.}~\bibnamefont
  {{Lasenby}}},\ }\href {\doibase 10.1086/309179} {\bibfield  {journal}
  {\bibinfo  {journal} {\apj}\ }\textbf {\bibinfo {volume} {538}},\ \bibinfo
  {pages} {473} (\bibinfo {year} {2000})},\ \Eprint
  {http://arxiv.org/abs/astro-ph/9911177} {astro-ph/9911177} \BibitemShut
  {NoStop}%
\bibitem [{\citenamefont {Hojjati}\ \emph {et~al.}(2011)\citenamefont
  {Hojjati}, \citenamefont {Pogosian},\ and\ \citenamefont
  {Zhao}}]{Hojjati:2011ix}%
  \BibitemOpen
  \bibfield  {author} {\bibinfo {author} {\bibfnamefont {A.}~\bibnamefont
  {Hojjati}}, \bibinfo {author} {\bibfnamefont {L.}~\bibnamefont {Pogosian}}, \
  and\ \bibinfo {author} {\bibfnamefont {G.-B.}\ \bibnamefont {Zhao}},\ }\href
  {\doibase 10.1088/1475-7516/2011/08/005} {\bibfield  {journal} {\bibinfo
  {journal} {JCAP}\ }\textbf {\bibinfo {volume} {1108}},\ \bibinfo {pages}
  {005} (\bibinfo {year} {2011})},\ \Eprint {http://arxiv.org/abs/1106.4543}
  {arXiv:1106.4543 [astro-ph.CO]} \BibitemShut {NoStop}%
\bibitem [{\citenamefont {Raveri}\ and\ \citenamefont
  {Hu}(2018)}]{Raveri:2018wln}%
  \BibitemOpen
  \bibfield  {author} {\bibinfo {author} {\bibfnamefont {M.}~\bibnamefont
  {Raveri}}\ and\ \bibinfo {author} {\bibfnamefont {W.}~\bibnamefont {Hu}},\
  }\href@noop {} {\  (\bibinfo {year} {2018})},\ \Eprint
  {http://arxiv.org/abs/1806.04649} {arXiv:1806.04649 [astro-ph.CO]}
  \BibitemShut {NoStop}%
\bibitem [{\citenamefont {Lorenz}\ \emph {et~al.}(2017)\citenamefont {Lorenz},
  \citenamefont {Calabrese},\ and\ \citenamefont {Alonso}}]{Lorenz:2017fgo}%
  \BibitemOpen
  \bibfield  {author} {\bibinfo {author} {\bibfnamefont {C.~S.}\ \bibnamefont
  {Lorenz}}, \bibinfo {author} {\bibfnamefont {E.}~\bibnamefont {Calabrese}}, \
  and\ \bibinfo {author} {\bibfnamefont {D.}~\bibnamefont {Alonso}},\ }\href
  {\doibase 10.1103/PhysRevD.96.043510} {\bibfield  {journal} {\bibinfo
  {journal} {Phys. Rev.}\ }\textbf {\bibinfo {volume} {D96}},\ \bibinfo {pages}
  {043510} (\bibinfo {year} {2017})},\ \Eprint
  {http://arxiv.org/abs/1706.00730} {arXiv:1706.00730 [astro-ph.CO]}
  \BibitemShut {NoStop}%
\bibitem [{\citenamefont {Ishak}(2018)}]{Ishak:2018his}%
  \BibitemOpen
  \bibfield  {author} {\bibinfo {author} {\bibfnamefont {M.}~\bibnamefont
  {Ishak}},\ }\href@noop {} {\  (\bibinfo {year} {2018})},\ \Eprint
  {http://arxiv.org/abs/1806.10122} {arXiv:1806.10122 [astro-ph.CO]}
  \BibitemShut {NoStop}%
\bibitem [{\citenamefont {Hu}\ and\ \citenamefont {Joyce}(2017)}]{Hu:2016wfa}%
  \BibitemOpen
  \bibfield  {author} {\bibinfo {author} {\bibfnamefont {W.}~\bibnamefont
  {Hu}}\ and\ \bibinfo {author} {\bibfnamefont {A.}~\bibnamefont {Joyce}},\
  }\href {\doibase 10.1103/PhysRevD.95.043529} {\bibfield  {journal} {\bibinfo
  {journal} {Phys. Rev.}\ }\textbf {\bibinfo {volume} {D95}},\ \bibinfo {pages}
  {043529} (\bibinfo {year} {2017})},\ \Eprint
  {http://arxiv.org/abs/1612.02454} {arXiv:1612.02454 [astro-ph.CO]}
  \BibitemShut {NoStop}%
\bibitem [{\citenamefont {{Hu}}\ and\ \citenamefont
  {{Sugiyama}}(1995)}]{Hu1995}%
  \BibitemOpen
  \bibfield  {author} {\bibinfo {author} {\bibfnamefont {W.}~\bibnamefont
  {{Hu}}}\ and\ \bibinfo {author} {\bibfnamefont {N.}~\bibnamefont
  {{Sugiyama}}},\ }\href {\doibase 10.1086/175624} {\bibfield  {journal}
  {\bibinfo  {journal} {\apj}\ }\textbf {\bibinfo {volume} {444}},\ \bibinfo
  {pages} {489} (\bibinfo {year} {1995})},\ \Eprint
  {http://arxiv.org/abs/astro-ph/9407093} {astro-ph/9407093} \BibitemShut
  {NoStop}%
\bibitem [{\citenamefont {Baumann}\ \emph {et~al.}(2016)\citenamefont
  {Baumann}, \citenamefont {Green}, \citenamefont {Meyers},\ and\ \citenamefont
  {Wallisch}}]{Baumann:2015rya}%
  \BibitemOpen
  \bibfield  {author} {\bibinfo {author} {\bibfnamefont {D.}~\bibnamefont
  {Baumann}}, \bibinfo {author} {\bibfnamefont {D.}~\bibnamefont {Green}},
  \bibinfo {author} {\bibfnamefont {J.}~\bibnamefont {Meyers}}, \ and\ \bibinfo
  {author} {\bibfnamefont {B.}~\bibnamefont {Wallisch}},\ }\href {\doibase
  10.1088/1475-7516/2016/01/007} {\bibfield  {journal} {\bibinfo  {journal}
  {JCAP}\ }\textbf {\bibinfo {volume} {1601}},\ \bibinfo {pages} {007}
  (\bibinfo {year} {2016})},\ \Eprint {http://arxiv.org/abs/1508.06342}
  {arXiv:1508.06342 [astro-ph.CO]} \BibitemShut {NoStop}%
\bibitem [{\citenamefont {{Hu}}\ and\ \citenamefont
  {{Sugiyama}}(1996)}]{Hu1996}%
  \BibitemOpen
  \bibfield  {author} {\bibinfo {author} {\bibfnamefont {W.}~\bibnamefont
  {{Hu}}}\ and\ \bibinfo {author} {\bibfnamefont {N.}~\bibnamefont
  {{Sugiyama}}},\ }\href {\doibase 10.1086/177989} {\bibfield  {journal}
  {\bibinfo  {journal} {\apj}\ }\textbf {\bibinfo {volume} {471}},\ \bibinfo
  {pages} {542} (\bibinfo {year} {1996})},\ \Eprint
  {http://arxiv.org/abs/astro-ph/9510117} {astro-ph/9510117} \BibitemShut
  {NoStop}%
\bibitem [{\citenamefont {Hu}\ and\ \citenamefont
  {Eisenstein}(1998)}]{Hu:1997vi}%
  \BibitemOpen
  \bibfield  {author} {\bibinfo {author} {\bibfnamefont {W.}~\bibnamefont
  {Hu}}\ and\ \bibinfo {author} {\bibfnamefont {D.~J.}\ \bibnamefont
  {Eisenstein}},\ }\href {\doibase 10.1086/305585} {\bibfield  {journal}
  {\bibinfo  {journal} {Astrophys. J.}\ }\textbf {\bibinfo {volume} {498}},\
  \bibinfo {pages} {497} (\bibinfo {year} {1998})},\ \Eprint
  {http://arxiv.org/abs/astro-ph/9710216} {arXiv:astro-ph/9710216 [astro-ph]}
  \BibitemShut {NoStop}%
\bibitem [{\citenamefont {Hui}\ and\ \citenamefont
  {Parfrey}(2008)}]{Hui:2007zh}%
  \BibitemOpen
  \bibfield  {author} {\bibinfo {author} {\bibfnamefont {L.}~\bibnamefont
  {Hui}}\ and\ \bibinfo {author} {\bibfnamefont {K.~P.}\ \bibnamefont
  {Parfrey}},\ }\href {\doibase 10.1103/PhysRevD.77.043527} {\bibfield
  {journal} {\bibinfo  {journal} {Phys. Rev.}\ }\textbf {\bibinfo {volume}
  {D77}},\ \bibinfo {pages} {043527} (\bibinfo {year} {2008})},\ \Eprint
  {http://arxiv.org/abs/0712.1162} {arXiv:0712.1162 [astro-ph]} \BibitemShut
  {NoStop}%
\bibitem [{\citenamefont {Chiang}\ \emph {et~al.}(2018)\citenamefont {Chiang},
  \citenamefont {Hu}, \citenamefont {Li},\ and\ \citenamefont
  {Loverde}}]{Chiang:2017vuk}%
  \BibitemOpen
  \bibfield  {author} {\bibinfo {author} {\bibfnamefont {C.-T.}\ \bibnamefont
  {Chiang}}, \bibinfo {author} {\bibfnamefont {W.}~\bibnamefont {Hu}}, \bibinfo
  {author} {\bibfnamefont {Y.}~\bibnamefont {Li}}, \ and\ \bibinfo {author}
  {\bibfnamefont {M.}~\bibnamefont {Loverde}},\ }\href {\doibase
  10.1103/PhysRevD.97.123526} {\bibfield  {journal} {\bibinfo  {journal} {Phys.
  Rev.}\ }\textbf {\bibinfo {volume} {D97}},\ \bibinfo {pages} {123526}
  (\bibinfo {year} {2018})},\ \Eprint {http://arxiv.org/abs/1710.01310}
  {arXiv:1710.01310 [astro-ph.CO]} \BibitemShut {NoStop}%
\bibitem [{\citenamefont {Long}\ \emph {et~al.}(2018)\citenamefont {Long},
  \citenamefont {Raveri}, \citenamefont {Hu},\ and\ \citenamefont
  {Dodelson}}]{Long:2017dru}%
  \BibitemOpen
  \bibfield  {author} {\bibinfo {author} {\bibfnamefont {A.~J.}\ \bibnamefont
  {Long}}, \bibinfo {author} {\bibfnamefont {M.}~\bibnamefont {Raveri}},
  \bibinfo {author} {\bibfnamefont {W.}~\bibnamefont {Hu}}, \ and\ \bibinfo
  {author} {\bibfnamefont {S.}~\bibnamefont {Dodelson}},\ }\href {\doibase
  10.1103/PhysRevD.97.043510} {\bibfield  {journal} {\bibinfo  {journal} {Phys.
  Rev.}\ }\textbf {\bibinfo {volume} {D97}},\ \bibinfo {pages} {043510}
  (\bibinfo {year} {2018})},\ \Eprint {http://arxiv.org/abs/1711.08434}
  {arXiv:1711.08434 [astro-ph.CO]} \BibitemShut {NoStop}%
\bibitem [{\citenamefont {Ade}\ \emph {et~al.}(2016{\natexlab{b}})\citenamefont
  {Ade} \emph {et~al.}}]{Ade:2015xua}%
  \BibitemOpen
  \bibfield  {author} {\bibinfo {author} {\bibfnamefont {P.~A.~R.}\
  \bibnamefont {Ade}} \emph {et~al.} (\bibinfo {collaboration} {Planck}),\
  }\href {\doibase 10.1051/0004-6361/201525830} {\bibfield  {journal} {\bibinfo
   {journal} {Astron. Astrophys.}\ }\textbf {\bibinfo {volume} {594}},\
  \bibinfo {pages} {A13} (\bibinfo {year} {2016}{\natexlab{b}})},\ \Eprint
  {http://arxiv.org/abs/1502.01589} {arXiv:1502.01589 [astro-ph.CO]}
  \BibitemShut {NoStop}%
\bibitem [{\citenamefont {Aghanim}\ \emph {et~al.}(2016)\citenamefont {Aghanim}
  \emph {et~al.}}]{Aghanim:2015xee}%
  \BibitemOpen
  \bibfield  {author} {\bibinfo {author} {\bibfnamefont {N.}~\bibnamefont
  {Aghanim}} \emph {et~al.} (\bibinfo {collaboration} {Planck}),\ }\href
  {\doibase 10.1051/0004-6361/201526926} {\bibfield  {journal} {\bibinfo
  {journal} {Astron. Astrophys.}\ }\textbf {\bibinfo {volume} {594}},\ \bibinfo
  {pages} {A11} (\bibinfo {year} {2016})},\ \Eprint
  {http://arxiv.org/abs/1507.02704} {arXiv:1507.02704 [astro-ph.CO]}
  \BibitemShut {NoStop}%
\bibitem [{\citenamefont {Ade}\ \emph {et~al.}(2016{\natexlab{c}})\citenamefont
  {Ade} \emph {et~al.}}]{Ade:2015zua}%
  \BibitemOpen
  \bibfield  {author} {\bibinfo {author} {\bibfnamefont {P.~A.~R.}\
  \bibnamefont {Ade}} \emph {et~al.} (\bibinfo {collaboration} {Planck}),\
  }\href {\doibase 10.1051/0004-6361/201525941} {\bibfield  {journal} {\bibinfo
   {journal} {Astron. Astrophys.}\ }\textbf {\bibinfo {volume} {594}},\
  \bibinfo {pages} {A15} (\bibinfo {year} {2016}{\natexlab{c}})},\ \Eprint
  {http://arxiv.org/abs/1502.01591} {arXiv:1502.01591 [astro-ph.CO]}
  \BibitemShut {NoStop}%
\bibitem [{\citenamefont {Riess}\ \emph {et~al.}(2016)\citenamefont {Riess}
  \emph {et~al.}}]{Riess:2016jrr}%
  \BibitemOpen
  \bibfield  {author} {\bibinfo {author} {\bibfnamefont {A.~G.}\ \bibnamefont
  {Riess}} \emph {et~al.},\ }\href {\doibase 10.3847/0004-637X/826/1/56}
  {\bibfield  {journal} {\bibinfo  {journal} {Astrophys. J.}\ }\textbf
  {\bibinfo {volume} {826}},\ \bibinfo {pages} {56} (\bibinfo {year} {2016})},\
  \Eprint {http://arxiv.org/abs/1604.01424} {arXiv:1604.01424 [astro-ph.CO]}
  \BibitemShut {NoStop}%
\bibitem [{\citenamefont {Heymans}\ \emph {et~al.}(2013)\citenamefont {Heymans}
  \emph {et~al.}}]{Heymans:2013fya}%
  \BibitemOpen
  \bibfield  {author} {\bibinfo {author} {\bibfnamefont {C.}~\bibnamefont
  {Heymans}} \emph {et~al.},\ }\href {\doibase 10.1093/mnras/stt601} {\bibfield
   {journal} {\bibinfo  {journal} {Mon. Not. Roy. Astron. Soc.}\ }\textbf
  {\bibinfo {volume} {432}},\ \bibinfo {pages} {2433} (\bibinfo {year}
  {2013})},\ \Eprint {http://arxiv.org/abs/1303.1808} {arXiv:1303.1808
  [astro-ph.CO]} \BibitemShut {NoStop}%
\bibitem [{\citenamefont {Joudaki}\ \emph
  {et~al.}(2017{\natexlab{b}})\citenamefont {Joudaki} \emph
  {et~al.}}]{Joudaki:2016mvz}%
  \BibitemOpen
  \bibfield  {author} {\bibinfo {author} {\bibfnamefont {S.}~\bibnamefont
  {Joudaki}} \emph {et~al.},\ }\href {\doibase 10.1093/mnras/stw2665}
  {\bibfield  {journal} {\bibinfo  {journal} {Mon. Not. Roy. Astron. Soc.}\
  }\textbf {\bibinfo {volume} {465}},\ \bibinfo {pages} {2033} (\bibinfo {year}
  {2017}{\natexlab{b}})},\ \Eprint {http://arxiv.org/abs/1601.05786}
  {arXiv:1601.05786 [astro-ph.CO]} \BibitemShut {NoStop}%
\bibitem [{\citenamefont {Ross}\ \emph {et~al.}(2015)\citenamefont {Ross},
  \citenamefont {Samushia}, \citenamefont {Howlett}, \citenamefont {Percival},
  \citenamefont {Burden},\ and\ \citenamefont {Manera}}]{Ross:2014qpa}%
  \BibitemOpen
  \bibfield  {author} {\bibinfo {author} {\bibfnamefont {A.~J.}\ \bibnamefont
  {Ross}}, \bibinfo {author} {\bibfnamefont {L.}~\bibnamefont {Samushia}},
  \bibinfo {author} {\bibfnamefont {C.}~\bibnamefont {Howlett}}, \bibinfo
  {author} {\bibfnamefont {W.~J.}\ \bibnamefont {Percival}}, \bibinfo {author}
  {\bibfnamefont {A.}~\bibnamefont {Burden}}, \ and\ \bibinfo {author}
  {\bibfnamefont {M.}~\bibnamefont {Manera}},\ }\href {\doibase
  10.1093/mnras/stv154} {\bibfield  {journal} {\bibinfo  {journal} {Mon. Not.
  Roy. Astron. Soc.}\ }\textbf {\bibinfo {volume} {449}},\ \bibinfo {pages}
  {835} (\bibinfo {year} {2015})},\ \Eprint {http://arxiv.org/abs/1409.3242}
  {arXiv:1409.3242 [astro-ph.CO]} \BibitemShut {NoStop}%
\bibitem [{\citenamefont {Beutler}\ \emph {et~al.}(2011)\citenamefont
  {Beutler}, \citenamefont {Blake}, \citenamefont {Colless}, \citenamefont
  {Jones}, \citenamefont {Staveley-Smith}, \citenamefont {Campbell},
  \citenamefont {Parker}, \citenamefont {Saunders},\ and\ \citenamefont
  {Watson}}]{Beutler:2011hx}%
  \BibitemOpen
  \bibfield  {author} {\bibinfo {author} {\bibfnamefont {F.}~\bibnamefont
  {Beutler}}, \bibinfo {author} {\bibfnamefont {C.}~\bibnamefont {Blake}},
  \bibinfo {author} {\bibfnamefont {M.}~\bibnamefont {Colless}}, \bibinfo
  {author} {\bibfnamefont {D.~H.}\ \bibnamefont {Jones}}, \bibinfo {author}
  {\bibfnamefont {L.}~\bibnamefont {Staveley-Smith}}, \bibinfo {author}
  {\bibfnamefont {L.}~\bibnamefont {Campbell}}, \bibinfo {author}
  {\bibfnamefont {Q.}~\bibnamefont {Parker}}, \bibinfo {author} {\bibfnamefont
  {W.}~\bibnamefont {Saunders}}, \ and\ \bibinfo {author} {\bibfnamefont
  {F.}~\bibnamefont {Watson}},\ }\href {\doibase
  10.1111/j.1365-2966.2011.19250.x} {\bibfield  {journal} {\bibinfo  {journal}
  {Mon. Not. Roy. Astron. Soc.}\ }\textbf {\bibinfo {volume} {416}},\ \bibinfo
  {pages} {3017} (\bibinfo {year} {2011})},\ \Eprint
  {http://arxiv.org/abs/1106.3366} {arXiv:1106.3366 [astro-ph.CO]} \BibitemShut
  {NoStop}%
\bibitem [{\citenamefont {Betoule}\ \emph {et~al.}(2014)\citenamefont {Betoule}
  \emph {et~al.}}]{Betoule:2014frx}%
  \BibitemOpen
  \bibfield  {author} {\bibinfo {author} {\bibfnamefont {M.}~\bibnamefont
  {Betoule}} \emph {et~al.} (\bibinfo {collaboration} {SDSS}),\ }\href
  {\doibase 10.1051/0004-6361/201423413} {\bibfield  {journal} {\bibinfo
  {journal} {Astron. Astrophys.}\ }\textbf {\bibinfo {volume} {568}},\ \bibinfo
  {pages} {A22} (\bibinfo {year} {2014})},\ \Eprint
  {http://arxiv.org/abs/1401.4064} {arXiv:1401.4064 [astro-ph.CO]} \BibitemShut
  {NoStop}%
\bibitem [{\citenamefont {{Addison}}\ \emph {et~al.}(2016)\citenamefont
  {{Addison}}, \citenamefont {{Huang}}, \citenamefont {{Watts}}, \citenamefont
  {{Bennett}}, \citenamefont {{Halpern}}, \citenamefont {{Hinshaw}},\ and\
  \citenamefont {{Weiland}}}]{Addison2016}%
  \BibitemOpen
  \bibfield  {author} {\bibinfo {author} {\bibfnamefont {G.~E.}\ \bibnamefont
  {{Addison}}}, \bibinfo {author} {\bibfnamefont {Y.}~\bibnamefont {{Huang}}},
  \bibinfo {author} {\bibfnamefont {D.~J.}\ \bibnamefont {{Watts}}}, \bibinfo
  {author} {\bibfnamefont {C.~L.}\ \bibnamefont {{Bennett}}}, \bibinfo {author}
  {\bibfnamefont {M.}~\bibnamefont {{Halpern}}}, \bibinfo {author}
  {\bibfnamefont {G.}~\bibnamefont {{Hinshaw}}}, \ and\ \bibinfo {author}
  {\bibfnamefont {J.~L.}\ \bibnamefont {{Weiland}}},\ }\href {\doibase
  10.3847/0004-637X/818/2/132} {\bibfield  {journal} {\bibinfo  {journal}
  {\apj}\ }\textbf {\bibinfo {volume} {818}},\ \bibinfo {eid} {132} (\bibinfo
  {year} {2016})},\ \Eprint {http://arxiv.org/abs/1511.00055}
  {arXiv:1511.00055} \BibitemShut {NoStop}%
\bibitem [{\citenamefont {Aghanim}\ \emph {et~al.}(2017)\citenamefont {Aghanim}
  \emph {et~al.}}]{Aghanim:2016sns}%
  \BibitemOpen
  \bibfield  {author} {\bibinfo {author} {\bibfnamefont {N.}~\bibnamefont
  {Aghanim}} \emph {et~al.} (\bibinfo {collaboration} {Planck}),\ }\href
  {\doibase 10.1051/0004-6361/201629504} {\bibfield  {journal} {\bibinfo
  {journal} {Astron. Astrophys.}\ }\textbf {\bibinfo {volume} {607}},\ \bibinfo
  {pages} {A95} (\bibinfo {year} {2017})},\ \Eprint
  {http://arxiv.org/abs/1608.02487} {arXiv:1608.02487 [astro-ph.CO]}
  \BibitemShut {NoStop}%
\bibitem [{\citenamefont {{Obied}}\ \emph {et~al.}(2017)\citenamefont
  {{Obied}}, \citenamefont {{Dvorkin}}, \citenamefont {{Heinrich}},
  \citenamefont {{Hu}},\ and\ \citenamefont {{Miranda}}}]{Obied2017}%
  \BibitemOpen
  \bibfield  {author} {\bibinfo {author} {\bibfnamefont {G.}~\bibnamefont
  {{Obied}}}, \bibinfo {author} {\bibfnamefont {C.}~\bibnamefont {{Dvorkin}}},
  \bibinfo {author} {\bibfnamefont {C.}~\bibnamefont {{Heinrich}}}, \bibinfo
  {author} {\bibfnamefont {W.}~\bibnamefont {{Hu}}}, \ and\ \bibinfo {author}
  {\bibfnamefont {V.}~\bibnamefont {{Miranda}}},\ }\href@noop {} {\bibfield
  {journal} {\bibinfo  {journal} {ArXiv e-prints}\ } (\bibinfo {year}
  {2017})},\ \Eprint {http://arxiv.org/abs/1706.09412} {arXiv:1706.09412}
  \BibitemShut {NoStop}%
\bibitem [{\citenamefont {Hu}\ and\ \citenamefont {Raveri}(2015)}]{Hu:2015rva}%
  \BibitemOpen
  \bibfield  {author} {\bibinfo {author} {\bibfnamefont {B.}~\bibnamefont
  {Hu}}\ and\ \bibinfo {author} {\bibfnamefont {M.}~\bibnamefont {Raveri}},\
  }\href {\doibase 10.1103/PhysRevD.91.123515} {\bibfield  {journal} {\bibinfo
  {journal} {Phys. Rev.}\ }\textbf {\bibinfo {volume} {D91}},\ \bibinfo {pages}
  {123515} (\bibinfo {year} {2015})},\ \Eprint
  {http://arxiv.org/abs/1502.06599} {arXiv:1502.06599 [astro-ph.CO]}
  \BibitemShut {NoStop}%
\bibitem [{\citenamefont {Motloch}\ and\ \citenamefont
  {Hu}(2018)}]{Motloch:2018pjy}%
  \BibitemOpen
  \bibfield  {author} {\bibinfo {author} {\bibfnamefont {P.}~\bibnamefont
  {Motloch}}\ and\ \bibinfo {author} {\bibfnamefont {W.}~\bibnamefont {Hu}},\
  }\href {\doibase 10.1103/PhysRevD.97.103536} {\bibfield  {journal} {\bibinfo
  {journal} {Phys. Rev.}\ }\textbf {\bibinfo {volume} {D97}},\ \bibinfo {pages}
  {103536} (\bibinfo {year} {2018})},\ \Eprint
  {http://arxiv.org/abs/1803.11526} {arXiv:1803.11526 [astro-ph.CO]}
  \BibitemShut {NoStop}%
\bibitem [{\citenamefont {{Ma}}\ and\ \citenamefont
  {{Bertschinger}}(1995)}]{Ma1995}%
  \BibitemOpen
  \bibfield  {author} {\bibinfo {author} {\bibfnamefont {C.-P.}\ \bibnamefont
  {{Ma}}}\ and\ \bibinfo {author} {\bibfnamefont {E.}~\bibnamefont
  {{Bertschinger}}},\ }\href {\doibase 10.1086/176550} {\bibfield  {journal}
  {\bibinfo  {journal} {\apj}\ }\textbf {\bibinfo {volume} {455}},\ \bibinfo
  {pages} {7} (\bibinfo {year} {1995})},\ \Eprint
  {http://arxiv.org/abs/astro-ph/9506072} {astro-ph/9506072} \BibitemShut
  {NoStop}%
\bibitem [{MGC()}]{MGCAMBPC}%
  \BibitemOpen
  \href@noop {} {}\bibinfo {howpublished} {Personal communication with the
  MGCAMB authors.}\BibitemShut {Stop}%
\end{thebibliography}%
\end{document}